%% file: Main.tex
\RequirePackage[10pt]{type1ec}

\documentclass[a4paper,longmktitle]{cas-sc}

\ExplSyntaxOn
\cs_if_exist:NF \vbox_unpack_clear:N
  { \cs_new_eq:NN \vbox_unpack_clear:N \vbox_unpack_drop:N }
\ExplSyntaxOff

\usepackage[numbers]{natbib}
\usepackage{booktabs}
\usepackage{tabularx}
\usepackage{threeparttable}
\usepackage{array}
\usepackage{graphicx}
\usepackage{float}
\usepackage{capt-of}
\usepackage[most]{tcolorbox}
\newcolumntype{Y}{>{\raggedright\arraybackslash}X}

\def\tsc#1{\csdef{#1}{\textsc{\lowercase{#1}}\xspace}}
\tsc{WGM}
\tsc{QE}
\begin{document}
\let\WriteBookmarks\relax
\def\floatpagepagefraction{1}
\def\textpagefraction{.001}

% Short title
\shorttitle{}    

% Short author
\shortauthors{}  

% Main title of the paper
\title [mode = title]{A Model-Agnostic Correction Framework for Short-Term Building Load Forecasting}

\author[a]{Shiqiao Zhou}
\credit{Writing -- original draft,
Writing -- review \& editing,
Resources,
Formal analysis,
Visualization,
Methodology,
Validation,
Software,
Investigation,
Data curation,
Conceptualization}

\author[a]{Huanbo Lyu}
\credit{Writing -- review \& editing,
Data curation,
Investigation}

\author[b]{Holger Schöner}
\credit{Writing -- review \& editing,
Supervision,
Conceptualization,
Funding acquisition}

\author[c]{Zipeng Wu}
\credit{Writing -- review \& editing}

\author[b]{Jessica Deuschel}
\credit{Conceptualization}

\author[a]{Freya Wentzel}
\credit{Writing -- review \& editing}

\author[d]{IAG Wilson}
\credit{Supervision,
Funding acquisition}

\author[a]{Shuo Wang}[orcid=0000-0003-1380-6428]
\cormark[1]
\ead{s.wang.2@bham.ac.uk}
\credit{Writing -- review \& editing,
Supervision,
Conceptualization,
Funding acquisition}

% ============================================================
% Affiliations
% ============================================================

\affiliation[a]{
    organization={School of Computer Science, University of Birmingham},
    addressline={Edgbaston},
    city={Birmingham},
    postcode={B15 2TT},
    state={West Midlands},
    country={United Kingdom}
}

\affiliation[b]{
    organization={Siemens AG},
    city={Munich},
    postcode={80333},
    state={Bavaria},
    country={Germany}
}

\affiliation[c]{
    organization={School of Mathematics, University of Birmingham},
    addressline={Edgbaston},
    city={Birmingham},
    postcode={B15 2TT},
    state={West Midlands},
    country={United Kingdom}
}

\affiliation[d]{
    organization={Birmingham Energy Institute, University of Birmingham},
    addressline={Edgbaston},
    city={Birmingham},
    postcode={B15 2TT},
    state={West Midlands},
    country={United Kingdom}
}

% ============================================================
% Corresponding author
% ============================================================

\cortext[cor1]{Corresponding author}

% Footnote text
% \fntext[1]{}

% For a title note without a number/mark
%\nonumnote{}

% Here goes the abstract
\begin{abstract}
Short-term building load forecasting remains challenging because load patterns can change rapidly with weather and operating conditions. Data-driven forecasting models may become less reliable when the current operating regime is rare or poorly represented during training, even when similar situations exist elsewhere in the historical record. Recent advances in retrieval-augmented time series forecasting offer a promising way to adapt predictions to changing conditions by directly reusing relevant historical cases at inference time. Motivated by this idea, we propose WeL-RAC, a
model-agnostic Weather-Load Retrieval-Augmented Correction framework
that explicitly retrieves such evidence at prediction time. WeL-RAC retrieves historical cases that are similar to the current condition based on both recent load states and horizon-aligned weather forecasts. Their observed future load trajectories are then aggregated and used to correct the initial prediction, without retraining or modifying the base forecaster. We evaluate WeL-RAC on 31 building cases across three datasets, eight forecasting backbones, and four horizons from 1 to 18 hours. The framework improves the corresponding base forecasters in 90 of 96 dataset-level comparisons. Beyond accuracy, success and failure case studies
further illustrate the framework's interpretability by
tracing how retrieved historical evidence improves or degrades
forecasts.
These findings support explicit historical retrieval as an effective
and interpretable complement to the building load
forecasting.
\end{abstract}

% Use if graphical abstract is present
%\begin{graphicalabstract}
%\includegraphics{}
%\end{graphicalabstract}

% Research highlights
% \begin{highlights}
% \item Historical evidence corrects building load forecasts without model retraining.
% \item Load-state summaries and weather semantics guide retrieval.
% \item Retrieval yields consistent forecasting gains across most building cases.
% \item Case studies reveal when retrieval correction improve or degrade forecasts.
% \end{highlights}

% Keywords
% Each keyword is seperated by \sep

\maketitle

% Main text
\input{sections/01-introduction-relatedwork}
\input{sections/02-methodology}

\input{sections/03-data-preparation-hyperparameter-search}
\input{sections/04-experiments-results}

\input{sections/05-retrieval-correction-analysis}
\input{sections/06-conclusion}

\medskip

% To print the credit authorship contribution details
% \printcredits

% \section*{Data Availability}
% Data will be made available on request.

% \section*{Declaration of competing interest}
% The authors declare that they have no known competing financial interests or personal relationships that could have appeared to influence the work reported in this paper.

% \section*{Declaration of generative AI and AI-assisted technologies in the manuscript preparation process}

% \noindent During the preparation of this work the author(s) used ChatGPT in order to improve the language and readability of the manuscript and to assist with code refinement. After using this tool/service, the author(s) reviewed and edited the content as needed and take(s) full responsibility for the content of the published article.

\section*{Acknowledgements}
The authors acknowledge the use of resources provided by the Isambard-AI National AI Research Resource (AIRR). Isambard-AI is operated by the University of Bristol and is funded by the UK Government’s Department for Science, Innovation and Technology (DSIT) via UK Research and Innovation; and the Science and Technology Facilities Council [ST/AIRR/I-A-I/1023]~\cite{mcintosh2024isambard}.

This research did not receive any specific grant from funding agencies in the public, commercial, or not-for-profit sectors.

%% Loading bibliography style file
% \bibliographystyle{model1-num-names}
\bibliographystyle{cas-model2-names}

% Loading bibliography database
\bibliography{cas-refs}

% Biography
%\bio{}
% Here goes the biography details.
%\endbio

%\bio{pic1}
% Here goes the biography details.
%\endbio

\end{document}

%% file: sections/01-introduction-relatedwork.tex
\section{Introduction}
% \subsection{Background and Motivation}
Buildings are a major source of energy demand and carbon emissions, making efficient building operation essential for the energy transition and climate-change mitigation \cite{IEA2025EnergyEfficiency}. In particular, short-term building energy forecasting supports operational decision-making by anticipating energy demand over the coming hours or days, thereby informing demand response, HVAC scheduling, peak-load management, model predictive control, and the optimisation of building energy efficiency and occupant comfort \cite{amasyali2018review, deb2017review, lyu2025data, li2025data}. Unlike long-term forecasts that primarily support planning and design, short-term forecasts must capture rapidly changing operating conditions and near-term external influences. This remains challenging because building loads are jointly affected by weather conditions, occupancy behaviour, HVAC operation, building thermal characteristics, and temporal usage patterns \cite{yu2022building, ni2024study, fan2026rethinking}. Their complex and time-varying interactions can induce nonlinear load dynamics and shifts in operating regimes, particularly under changing weather or abnormal building operation \cite{li2023large}.

Existing studies have developed a wide range of building load forecasting models, including statistical, machine learning, deep learning, and hybrid approaches \cite{deb2017review,zhang2021review}. These methods learn nonlinear temporal dependencies from historical loads and exogenous variables during training. Nevertheless, this training paradigm has a limitation in short-term load forecasting, where load patterns can change rapidly with weather and building operating conditions. Once trained, most parametric forecasters encode historical observations primarily in fixed model parameters and do not explicitly revisit individual past cases during inference \cite{han2025raft}. Consequently, their predictions may become less reliable when the current operating regime is rare or underrepresented in the training distribution, even when comparable segments exist somewhere in the historical record. Although continual or online adaptation can update forecasting models as operating conditions evolve \cite{li2023large}, a complementary approach is to access relevant historical experience directly at inference time without modifying the forecaster's parameters. When the recent load state and weather resemble those of a previously observed segment, its subsequent load trajectory can provide a useful historical reference for the current forecast \cite{jiang2023nsdar,genov2026adaptive}.

The value of such historical analogues has been recognized in similar-day and nearest-neighbor-based forecasting methods, which typically select historical days with similar calendar or weather conditions and use their load profiles to construct forecasts. Existing methods span different strategies for similar-day selection, forecasting, and their integration~\cite{alessandrini2015analog, ma2017energy, zhang2022similarity, jiang2023nsdar, genov2026adaptive}. However, similarity in building loads is inherently context dependent. Historical periods with similar recent load patterns may evolve differently under different future weather conditions, while weather-similar periods may correspond to different building operating states. Consequently, retrieval based on a single similarity criterion may select historical profiles whose future trajectories are not fully compatible with the current forecasting context. Moreover, because retrieved analogues represent only approximate matches to the current condition, using them as standalone forecasts may not be sufficiently reliable to replace an existing parametric forecaster.

\begin{figure}
  \centering
  \includegraphics[width=\linewidth, trim=0 0 0 0, clip]{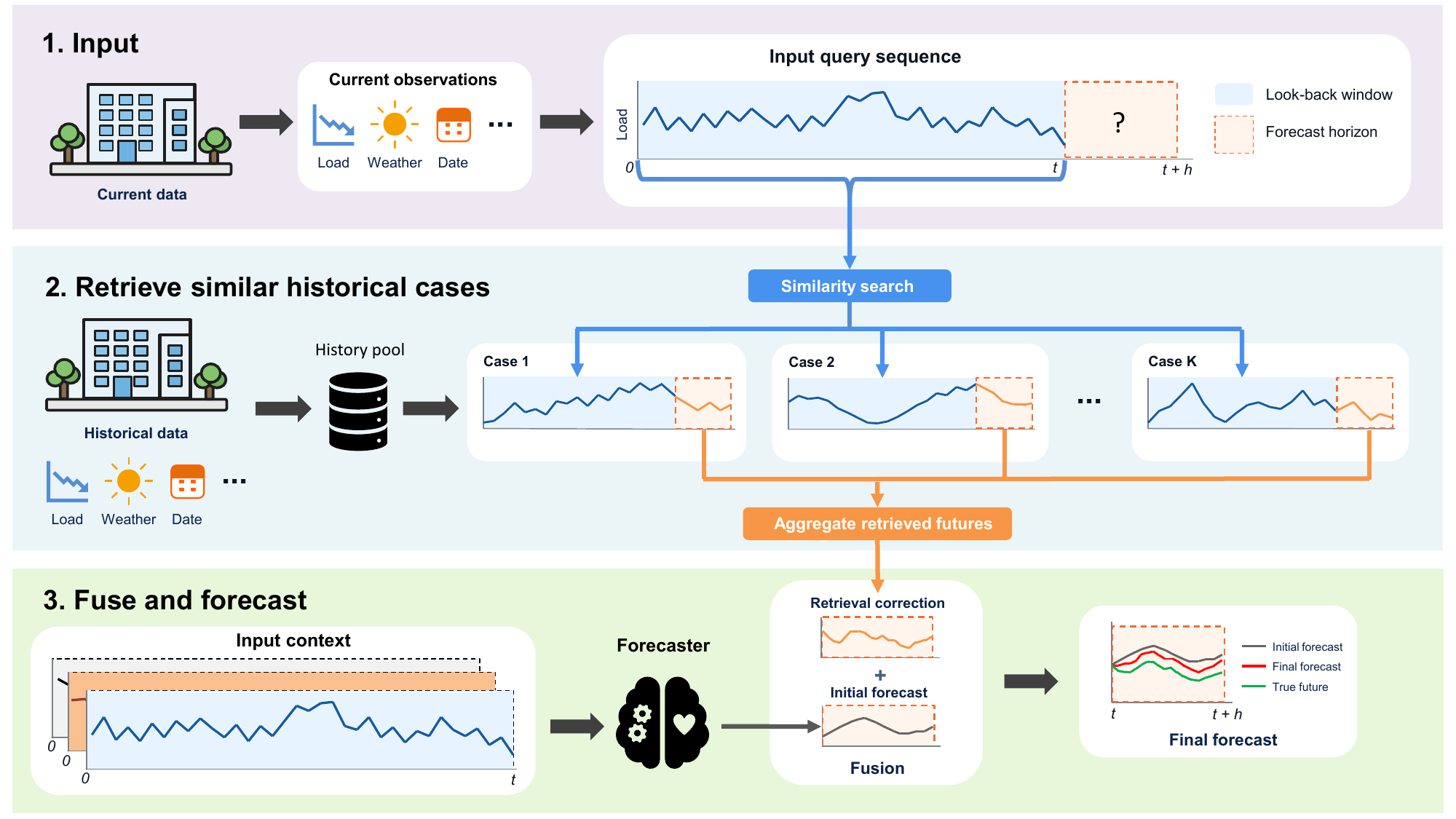}
  \caption{Illustration of retrieval-augmented forecasting. Taking building energy forecasting as an example, the process consists of three main steps: (1) the current load, weather, and other information are used to form the forecasting query; (2) similar historical cases are retrieved from a historical data pool based on the current input conditions, and their observed future load trajectories are aggregated as retrieval evidence; and (3) the retrieved evidence is combined with the initial prediction from the forecasting model to produce the final forecast.} 
  \label{fig:intro}
\end{figure}

Retrieval-augmented forecasting provides a natural way to complement a trained parametric forecaster with explicitly accessible historical information. Inspired by retrieval-augmented generation in natural language processing \cite{lewis2020retrieval}, recent time-series studies retrieve relevant historical segments and exploit their associated future trajectories or representations to improve forecasting \cite{yang2025timerag,han2025raft,zhou2026stationarity,cao2026enhancing}. As illustrated in Fig.~\ref{fig:intro}, a key idea is to reuse the observed futures of similar historical segments and combine them with the prediction of the forecasting model. This paradigm is particularly attractive for short-term building load forecasting, where predictions must respond rapidly to changing operating and weather conditions. However, many existing retrieval methods primarily determine relevance from the observed input sequence itself~\cite{yang2025timerag,han2025raft}. In usual, short-term building load forecasting not only consider the past load but also uses an additional source of information that can be exploited at retrieval time: weather forecasts over the prediction horizon are often available before the future load is observed. In particular, jointly considering the recent load state and weather conditions can help identify historical trajectories that are more compatible with the current forecasting context. Their observed future loads can then serve as a complementary reference for correcting the initial building load prediction. Such a retrieve-and-correct formulation enables fast adaptation to the current input while preserving the architecture and parameters of the forecaster.

Motivated by these considerations, this study proposes WeL-RAC, a \textbf{We}ather--\textbf{L}oad \textbf{R}etrieval-\textbf{A}ugmented \textbf{C}orrection framework for short-term building load forecasting. WeL-RAC combines two complementary retrieval views: a numerical representation of the recent load state and a semantic representation of the horizon-aligned weather forecast. The forecasting branch first generates an initial prediction using a base forecaster. The retrieval branch compares the current load state with historical load states in a numerical feature space, while comparing a textual summary of the upcoming weather forecast with horizon-aligned historical weather summaries in a semantic embedding space. The two similarity signals are then fused to retrieve the top-$K$ historical cases, whose observed future load trajectories are aggregated into a retrieved future trajectory. Finally, the correction module performs weighted fusion between the initial forecast and the retrieved future trajectory to produce the final prediction.

The main contributions of this study are summarized as follows:

\begin{itemize}
\item We formulate short-term building load forecasting as a retrieval-augmented correction problem and propose a model-agnostic framework that can enhance existing forecasters without modifying their internal architectures.

\item We develop a dual-view retrieval mechanism that jointly considers recent load states and weather contexts to identify relevant historical analogues and retrieve their corresponding future load trajectories without additional training of the retrieval module.

\item We provide case-based interpretability by explicitly linking each forecast correction to retrieved historical cases, allowing the source and direction of beneficial or harmful adjustments to be directly inspected.

\item We conduct extensive experiments on 31 building cases across three datasets, eight forecasting backbones, and four forecasting horizons, together with ablation, sensitivity, and computational cost analyses to systematically evaluate the proposed framework.

\end{itemize}

The remainder of this paper is organized as follows. Section~\ref{sec:RelatedWorks} reviews related work on time series forecasting and building load forecasting. Section~\ref{sec:methodology} presents the proposed WeL-RAC framework, including the forecasting, retrieval, and correction components. Section~\ref{sec:experimental_setup} describes the datasets, experimental settings, baselines, evaluation metrics, and hyperparameter selection. Section~\ref{sec:results} reports the main forecasting results together with ablation, sensitivity, and computational-cost analyses. Section~\ref{sec:case_study} provides case studies to further illustrate when retrieval correction succeeds or fails. Finally, Section~\ref{sec:conclusion} discusses the limitations of the proposed approach and concludes the paper.

\section{Related Works}
\label{sec:RelatedWorks}
\subsection{Time Series Forecasting}
% traditional/machine learning/LLM-based/Retrieval-based

Time series forecasting has progressed from classical statistical approaches, including exponential smoothing, autoregressive integrated moving average models, and state-space models, to machine-learning methods such as support vector regression, tree-based ensembles, and artificial neural networks~\cite{box2015time,hyndman2021forecasting}. Deep learning subsequently improved temporal representation learning through recurrent architectures such as long short-term memory networks~\cite{hochreiter1997long} and Transformer-based forecasters such as Informer and PatchTST~\cite{zhou2021informer,nie2023patchtst}. More recently, LLM-based and foundation-model approaches have explored numerical sequence tokenization, language-model reprogramming, and large-scale time series pre-training. Representative methods include LLMTime, Time-LLM, TimesFM and Chronos~\cite{gruver2023large,jin2024time,das2023decoder, ansari2024chronos}. Although these developments have substantially improved nonlinear modeling, long-range dependency learning, and cross-domain generalization, most existing approaches remain predominantly parametric: knowledge extracted from the training data is compressed into model parameters, while inference generally relies on a bounded observation window. Consequently, sparse but informative historical operating regimes may not be explicitly accessible when forecasts are generated. This limitation is particularly restrictive for building energy consumption forecasting, where similar load trajectories can evolve differently under different weather and operational conditions.

Retrieval-based forecasting provides a complementary non-parametric mechanism by explicitly accessing relevant historical cases at inference time. 
% Rooted in analog forecasting, nearest-neighbor regression, dynamic time warping, and case-based reasoning, 
Such ideas have roots in analog forecasting, nearest-neighbor regression, dynamic time warping, and case-based reasoning. Early analog-ensemble methods, for example, constructed probabilistic weather forecasts from observations associated with historical numerical weather predictions similar to the current forecast~\cite{delle2013probabilistic}, and later extended this principle to short-term wind-power forecasting using historical power observations associated with analogous weather forecasts~\cite{alessandrini2015novel}.
Modern retrieval-augmented forecasting transfers the retrieval-augmented generation principle from natural language processing to temporal prediction~\cite{lewis2020rag}. 
RAFT retrieves training windows similar to the current input and uses their corresponding future trajectories to augment forecasting~\cite{han2025raft}. Retrieval has also been combined with LLMs and time series foundation models: TimeRAG uses dynamic time warping to retrieve reference sequences for prompt augmentation, TimeRAF employs an end-to-end learnable retriever and channel prompting, and TS-RAG retrieves semantically relevant temporal segments and adaptively fuses them with foundation-model representations~\cite{yang2025timerag,11031238,ning2026ts}. More recent studies have further examined how retrieved evidence can be made more informative and reliable. SARAF balances retrieval relevance and regime diversity and performs stationarity-aware aggregation of retrieved futures~\cite{zhou2026stationarity}, while SERAF shows that numerical time-series retrieval and semantic retrieval can provide complementary historical evidence~\cite{zhou2026seraf}. Motivated by this complementary dual-view retrieval principle, WeL-RAC adapts and reformulates the two retrieval views for short-term building energy forecasting. Rather than matching raw historical input sequences, the numerical branch represents recent building demand using a compact load-state summary that summarizes the current operating condition. Meanwhile, the semantic branch moves beyond descriptions derived from the observed historical input and instead exploits horizon-aligned weather forecasts, retrieving historical cases under similar anticipated weather conditions. The resulting load-state and weather similarities are jointly used to identify historical analogues. Importantly, the retrieval and fusion process is training-free and operates as an external post-prediction correction layer, decoupled from the internal architecture of the forecasting model. WeL-RAC can therefore be attached to different forecasters without retraining or modifying their architectures, providing a lightweight and interpretable retrieval-augmented mechanism tailored to building energy forecasting.

\subsection{Building Energy Forecasting}

% Building energy forecasting supports applications such as HVAC control, demand response, peak-load management, and energy-efficient building operation. Compared with generic time series forecasting, building loads are strongly influenced by external and operational factors, including weather conditions, calendar patterns, occupancy behaviour, and HVAC operation \cite{luo2021forecasting}. Consequently, data-driven forecasting methods commonly incorporate historical energy consumption together with weather and calendar variables to capture nonlinear and context-dependent demand patterns \cite{zhao2012review,deb2017review,daut2017building, sun2020review}. A wide range of machine-learning and deep-learning approaches have been investigated for this purpose, while relatively simple models such as SVR and tree-based methods can remain competitive in building-level forecasting \cite{dong2005applying,fan2014development,fan2017short,wang2020building}.

Building load forecasting supports applications such as HVAC control, demand response, peak-load management, and energy-efficient building operation. Building loads are strongly influenced by external and operational factors, including weather conditions, calendar patterns, occupancy behaviour, and HVAC operation \cite{luo2021forecasting}.
Building energy and thermal modeling broadly follows physics-based and data-driven approaches, with the former exemplified by building simulation tools such as EnergyPlus \cite{crawley2001energyplus}. Hybrid approaches combine physical knowledge with data, as in resistance--capacitance grey-box models and physically consistent neural networks \cite{bacher2011identifying,di2022physically}.
Data-driven forecasting methods commonly incorporate historical energy consumption together with weather and calendar variables to capture nonlinear and context-dependent demand patterns \cite{zhao2012review,deb2017review,daut2017building,sun2020review}. A wide range of machine-learning and deep-learning approaches have been investigated for this purpose, while relatively simple models such as SVR and tree-based methods can remain competitive in building-level forecasting \cite{dong2005applying,fan2014development,fan2017short,wang2020building}.

Beyond directly incorporating external variables into general energy forecasting models, another line of research exploits historically similar operating conditions. Similar-day and analogue-based methods identify historical periods that resemble the current forecasting context according to temporal patterns, weather conditions, calendar information, or load characteristics. Earlier studies used such similarity to select more relevant historical samples or construct forecasts from comparable load profiles \cite{paudel2015support,ma2017energy,zhang2022similarity}. More recent work has further introduced learned similar-day screening and adaptive adjustment of historical analogue profiles under changing operating conditions \cite{jiang2023nsdar,genov2026adaptive}. These studies demonstrate that historically similar cases can provide useful empirical evidence for building energy forecasting, and can be viewed as important predecessors of retrieval-based forecasting.

However, existing similar-day and analogue-based approaches are typically designed as standalone forecasting strategies or training-data selection procedures. Historical analogues are therefore rarely formulated as an external, model-agnostic source of evidence that can directly correct predictions from heterogeneous forecasting models. Moreover, prior work suggests that both load conditions and weather conditions are important for identifying informative historical analogues~\cite{ma2017energy,zhang2022similarity}. This observation motivates the dual-view similarity design adopted in our retrieval-augmented correction framework, which jointly considers recent load states and future weather contexts when retrieving historical evidence for prediction correction.

%% file: sections/02-methodology.tex
\section{Methodology}
\label{sec:methodology}

This section introduces WeL-RAC, a \textbf{We}ather-\textbf{L}oad \textbf{R}etrieval-\textbf{A}ugmented \textbf{C}orrection framework for short-term building load forecasting. WeL-RAC separates parametric forecasting from non-parametric evidence retrieval. A trained base forecaster first produces an initial load prediction from the historical load and covariate window. The retrieval module then searches a training-period memory for historical cases that are similar to the current forecast origin in two complementary aspects: recent load state and horizon-aligned weather regime. The retrieved future load trajectories are aggregated as empirical evidence and fused with the initial prediction through a post-prediction correction module. This design allows retrieval to act as a lightweight calibration layer that can be attached to different forecasting backbones without changing their internal architectures.

\subsection{Framework Overview}
\label{subsec:framework_overview}

\begin{figure}
  \centering
  \includegraphics[width=\linewidth, trim=0 60 0 0, clip]{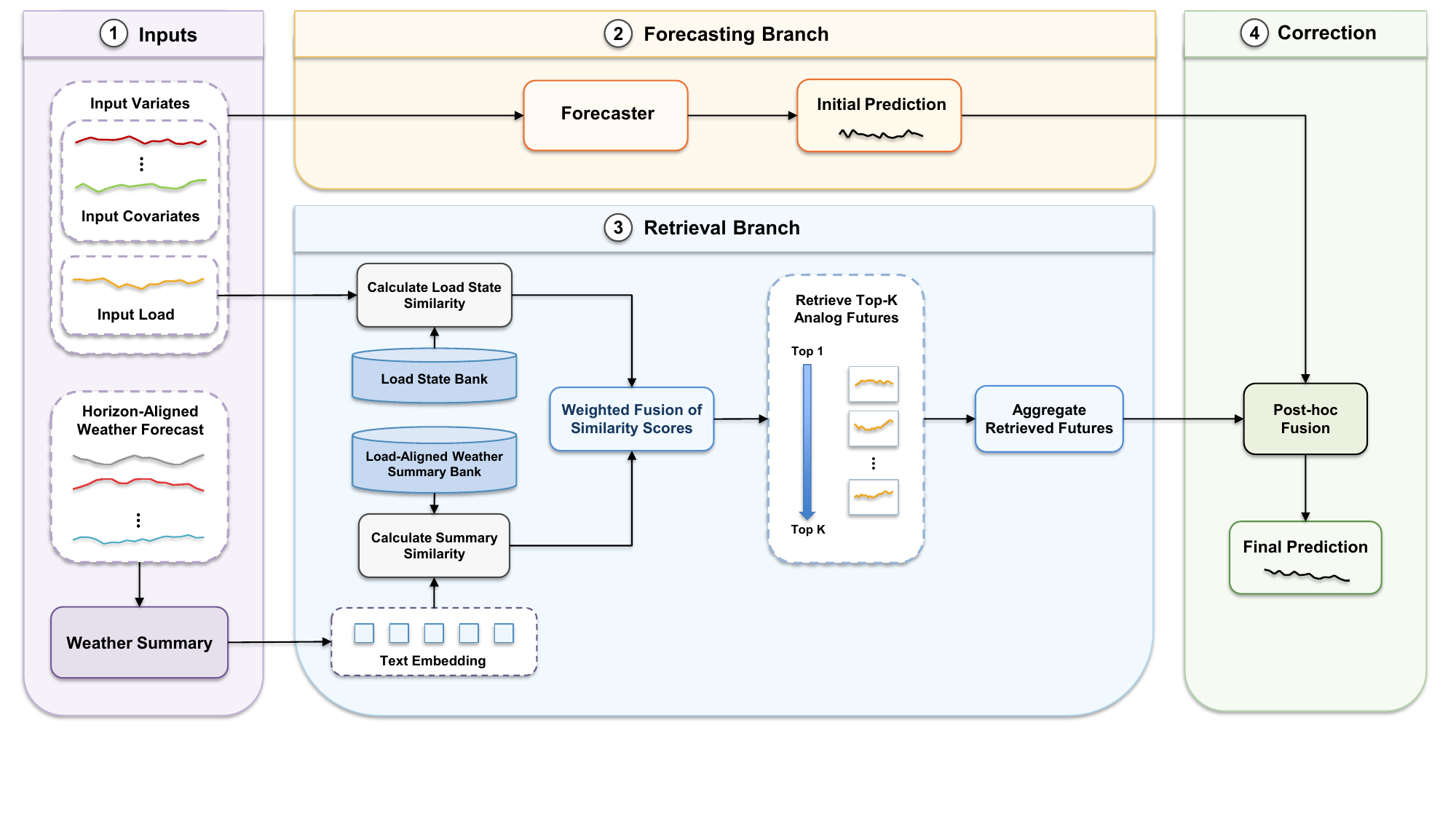}
  \caption{Schematic overview of WeL-RAC. The forecasting branch generates an initial prediction from historical load and covariates. The retrieval branch matches the current load state and horizon-aligned weather summary against historical memory banks, retrieves top-$K$ analogue futures, and aggregates their realised trajectories. The correction module performs post-prediction fusion between the initial prediction and the retrieved evidence to produce the final prediction.}
  \label{fig:method}
\end{figure}

Fig.~\ref{fig:method} shows the overall pipeline. At each forecast origin, the input component provides three aligned information sources: a historical load-covariate window for the base forecaster, a recent load window for load-state matching, and a horizon-aligned weather forecast for weather-summary retrieval. The forecasting branch maps the historical load-covariate window to an initial prediction. The retrieval branch compares the current query with two training-period banks: a load-state bank and a load-aligned weather-summary bank. The two similarity scores are fused into a single analogue score, from which the top-$K$ historical analogue futures are selected and aggregated. The correction module then combines the initial prediction with the aggregated retrieved evidence to generate the final forecast.

% The framework is organised around the following operations:
% \begin{enumerate}
%     \item a model-agnostic forecasting branch produces an anchor future load trajectory from historical load and covariate features;
%     \item a load-state bank stores normalised summarys of recent historical load windows;
%     \item a load-aligned weather-summary bank stores semantic embeddings of weather summaries generated from horizon-aligned weather windows;
%     \item the current query is represented by a recent load-state summary and a weather summary generated from the horizon-aligned weather forecast;
%     \item load-state similarity and weather-summary similarity are computed separately and fused through a weighted analogue score;
%     \item top-$K$ historical analogues provide their realised future load trajectories as non-parametric evidence;
%     \item the final prediction is obtained by a post-prediction fusion of the anchor trajectory and the retrieved evidence.
% \end{enumerate}

\subsection{Problem Definition and Formulation}
\label{subsec:problem_formulation}

\noindent\textbf{Definition 1 (Single building load forecasting).}
We consider the standard exogenous time series forecasting setting for building electricity consumption, where a multivariate input sequence is observed but only one target channel is predicted. For a single building, let
\begin{equation}
    \mathbf{x}_t =
    [y_t,\mathbf{c}_t^\top]^\top \in \mathbb{R}^{D}
\end{equation}
denote the observation at time step $t$, where $y_t \in \mathbb{R}$ is the target building load, also referred to as the single output target (OT), and $\mathbf{c}_t \in \mathbb{R}^{D-1}$ contains exogenous covariates such as calendar indicators, weather variables, and other available auxiliary features. For a forecast origin $t$, an input length $L$, and a prediction horizon $H$, the historical input window is
\begin{equation}
    \mathbf{X}
    =
    [\mathbf{x}_{t-L+1},\ldots,\mathbf{x}_{t}]
    \in \mathbb{R}^{L \times D}.
\end{equation}
The future target trajectory is
\begin{equation}
    \mathbf{y}
    =
    [y_{t+1},y_{t+2},\ldots,y_{t+H}]^\top
    \in \mathbb{R}^{H}.
\end{equation}
% For notational clarity, once the forecast origin $t$ and horizon $H$ are fixed, we omit them from most subsequent symbols and use $\mathbf{X}$, $\mathbf{W}^{\mathrm{q}}$, and $\mathbf{y}$ for the current forecasting instance.
For datasets containing multiple buildings, this formulation is applied independently to each building. WeL-RAC does not assume or model cross-building spatial dependencies.

\noindent\textbf{Definition 2 (Retrieval query).}
WeL-RAC uses two query signals for retrieval: a weather-summary query constructed from the horizon-aligned weather forecast and a recent load-state summary. Let $\mathbf{w}_t \in \mathbb{R}^{D_w}$ denote the weather variables used by the retrieval branch, such as temperature, humidity, dew point, precipitation, and wind speed when available. At inference time, the forecast weather window and its textual query are
\begin{equation}
    \mathbf{W}^{\mathrm{q}}
    =
    [\mathbf{w}^{\mathrm{fcst}}_{t+1},\ldots,\mathbf{w}^{\mathrm{fcst}}_{t+H}]
    \in \mathbb{R}^{H \times D_w},
    \qquad
    \mathbf{z}^{\mathrm{q}}
    =
    g(\mathbf{W}^{\mathrm{q}}).
\end{equation}
where the superscripts $\mathrm{q}$ and $\mathrm{h}$ denote query-side and historical-memory retrieval quantities, respectively. The operator $g(\cdot)$ is the shared weather-summary operator used throughout WeL-RAC. It maps both the current forecast-weather window and the historical horizon-aligned weather windows stored in memory to template-based natural-language descriptions under the same field-selection rules.
The input load segment is encoded into the query load-state summary as
\begin{equation}
    \mathbf{y}^{\mathrm{past}}_t
    =
    [y_{t-L+1},\ldots,y_t]^\top,
    \qquad
    \mathbf{s}^{\mathrm{q}}
    =
    \phi_s(\mathbf{y}^{\mathrm{past}}_t)
    \in \mathbb{R}^{d_s}.
\end{equation}
Here $\phi_s(\cdot)$ is a deterministic load-state encoder that summarises recent load level, variability, extrema, net change, trend, and short-term differences. The summary is standardised using training-set statistics and then $\ell_2$-normalised, yielding $\mathbf{s}^{\mathrm{q}}$. In the following derivations, $\mathbf{z}^{\mathrm{q}}$ and $\mathbf{s}^{\mathrm{q}}$ denote the weather-summary and load-state query for the current forecast origin.

\noindent\textbf{Definition 3 (Retrieval-corrected forecasting task).}
WeL-RAC is designed as a correction framework rather than a replacement for the base forecaster. Given the historical input $\mathbf{X}$, a trained base model first produces an initial prediction:
\begin{equation}
    \widehat{\mathbf{y}}^{\mathrm{initial}}
    =
    f_{\theta}(\mathbf{X}),
    \label{eq:base_forecast}
\end{equation}
where $\widehat{\mathbf{y}}^{\mathrm{initial}}\in\mathbb{R}^{H}$ is the initial prediction. The retrieval module $R(\cdot)$ then searches a horizon-specific training memory $\mathcal{M}$ using the query $(\mathbf{z}^{\mathrm{q}},\mathbf{s}^{\mathrm{q}})$ and returns an aggregated analogue future:
\begin{equation}
    \widetilde{\mathbf{y}}
    =
    R(\mathbf{z}^{\mathrm{q}},\mathbf{s}^{\mathrm{q}};\mathcal{M}).
\end{equation}
The final prediction is obtained by correcting the initial prediction with the retrieved evidence:
\begin{equation}
    \widehat{\mathbf{y}}
    =
    C(\widehat{\mathbf{y}}^{\mathrm{initial}},\widetilde{\mathbf{y}}).
\end{equation}
Here $C(\cdot)$ denotes the post-prediction correction module that fuses the initial prediction with the retrieved analogue evidence. The memory $\mathcal{M}$ is built only from the training period, so no validation or test future information is used for retrieval. This formulation makes the role of WeL-RAC explicit: the base model provides the initial prediction, while retrieval supplies instance-specific evidence for post-prediction correction.

\subsection{Forecasting Branch}
\label{subsec:base_forecasting_branch}

The forecasting branch produces the initial prediction to be corrected. The base forecaster $f_{\theta}$ may be instantiated as a linear temporal model, a tree-based regressor, a neural sequence model, a Transformer-based architecture, or a pretrained time series foundation model. It maps the historical load and covariates window $\mathbf{X}$ to the initial prediction $\widehat{\mathbf{y}}^{\mathrm{initial}}$ in Eq.~\eqref{eq:base_forecast}. WeL-RAC does not impose any architectural constraint on $f_{\theta}$ and does not modify its training objective. The retrieval module is therefore decoupled from the internal representation of the base model and operates as an external correction layer.

\subsection{Weather Summary Construction}
\label{subsec:weather_summary_construction}

The weather-summary channel represents any horizon-aligned weather window as a textual description before semantic embedding. A single deterministic operator $g(\cdot)$ is used on both sides of retrieval. For the current forecast origin, the future weather forecast $\mathbf{W}^{\mathrm{q}}$ is transformed into the query summary $\mathbf{z}^{\mathrm{q}}=g(\mathbf{W}^{\mathrm{q}})$. For a historical memory item $j$, the stored horizon-aligned weather window for the same lead-time range is transformed by the same operator:
\begin{equation}
    \mathbf{z}^{\mathrm{h}}_j
    =
    g(\mathbf{W}^{\mathrm{h}}_j),
\end{equation}
where $\mathbf{W}^{\mathrm{h}}_j$ is the historical horizon-aligned weather window stored in memory item $j$. Its dataset-specific source is described in Sec.~\ref{subsec:datasets_preprocessing}; in all cases, the same template, field-selection rules, and phrase categories are used for query and memory summaries. This ensures that weather contexts are embedded in a comparable textual space. The template is conditional: temperature and time fields are used whenever available; humidity and dew-point clauses are included when moisture variables exist; and precipitation or wind clauses are omitted for datasets that do not provide those variables. Fig.~\ref{fig:weather_template} gives the summary template, and Table~\ref{tab:weather_phrase_categories} lists the fields used to instantiate it.

% \begin{tcolorbox}[
%   top=1pt,
%   bottom=1pt,
%   left=1pt,
%   right=0.5pt,
%   colback=white,
%   colframe=black,
%   coltitle=white,
%   colbacktitle=black,
%   adjusted title={\textbf{\large Weather Summary Template:}}
% ]
% \small\itshape
% ``Weather summary for a \{horizon\}-hour window starting \{time\}.''\\
% ``The overall temperature pattern is \{temp\_mean\}, with a peak in the \{temp\_max\} range and a low in the \{temp\_min\} range.''\\
% ``The temperature trend is \{trend\}, with \{hot\_ratio\} hot-hour coverage.''\\
% ``Humidity feels \{humidity\}, and dew-point conditions are \{dewpoint\}.''\\
% ``Precipitation conditions are \{precip\}, with \{wind\} winds.''\\
% ``Humid-heat risk is \{risk\}.''
% \end{tcolorbox}
% \captionof{figure}{Conditional weather-summary template for representing a horizon-aligned weather window. Fields whose source variables are unavailable are omitted consistently in both query and memory summaries.}
% \label{fig:weather_template}

\begin{figure}
\centering
\begin{tcolorbox}[
  top=1pt,
  bottom=1pt,
  left=1pt,
  right=0.5pt,
  colback=white,
  colframe=black,
  coltitle=white,
  colbacktitle=black,
  adjusted title={\textbf{\large Weather Summary Template:}}
]
\small\itshape
``Weather summary for a \{horizon\}-hour window starting \{time\}.''\\
``The overall temperature pattern is \{temp\_mean\}, with a peak in the \{temp\_max\} range and a low in the \{temp\_min\} range.''\\
``The temperature trend is \{trend\}, with \{hot\_ratio\} hot-hour coverage.''\\
``Humidity feels \{humidity\}, and dew-point conditions are \{dewpoint\}.''\\
``Precipitation conditions are \{precip\}, with \{wind\} winds.''\\
``Humid-heat risk is \{risk\}.''
\end{tcolorbox}
\caption{
Conditional weather-summary template for representing both horizon-aligned forecast and historical weather windows. The template fields are defined and selected in Table~\ref{tab:weather_phrase_categories}.
}
\label{fig:weather_template}
\end{figure}

\begin{table}[H]
\begingroup
\centering
\caption{Weather-summary fields and phrase-selection rules.}
\label{tab:weather_phrase_categories}
\footnotesize
\renewcommand{\arraystretch}{1.08}
\begin{tabularx}{\linewidth}{@{}p{0.22\linewidth}p{0.30\linewidth}X@{}}
\toprule
\textbf{Template field} & \textbf{Source statistic} & \textbf{Phrase selection} \\
\midrule
\texttt{horizon} 
& Prediction length $H$ 
& Numeric horizon, expressed in hours for hourly datasets. \\

\texttt{time} 
& Start timestamp of the forecast window 
& night; morning; afternoon; evening. \\

\texttt{temp\_mean} 
& Mean temperature over the horizon 
& cold; mild; warm; hot; very hot. \\

\texttt{temp\_max} 
& Maximum temperature over the horizon 
& cool; mild; warm; hot; very hot. \\

\texttt{temp\_min} 
& Minimum temperature over the horizon 
& cold; cool; mild; warm; hot. \\

\texttt{trend} 
& Temperature change 
& falling quickly; falling; mostly steady; rising; rising quickly. \\

\texttt{hot\_ratio} 
& Fraction of horizon above the hot-temperature threshold 
& little to no; some; roughly half; mostly. \\

\texttt{humidity} 
& Mean relative humidity, if available 
& very dry; dry; moderate; humid; very humid. \\

\texttt{dewpoint} 
& Mean dew point, if available 
& very dry; dry; comfortable; sticky; tropical. \\

\texttt{precip} 
& Precipitation amount or precipitation flag, if available 
& dry; slight chance of rain; light rain; wet conditions. \\

\texttt{wind} 
& Mean or peak wind speed, if available 
& calm; light; moderate; windy. \\

\texttt{risk} 
& Joint hot-humid condition derived from temperature and moisture fields 
& low; elevated. \\
\bottomrule
\end{tabularx}
\par
\endgroup
\end{table}

A frozen text encoder $E(\cdot)$ maps each summary to an $\ell_2$-normalised embedding:
\begin{equation}
    \mathbf{e}^{\mathrm{q}}
    =
    \frac{E(\mathbf{z}^{\mathrm{q}})}{\lVert E(\mathbf{z}^{\mathrm{q}}) \rVert_2},
    \qquad
    \mathbf{e}^{\mathrm{h}}_j
    =
    \frac{E(\mathbf{z}^{\mathrm{h}}_j)}{\lVert E(\mathbf{z}^{\mathrm{h}}_j) \rVert_2}.
\end{equation}
This semantic representation supports retrieval over weather regimes rather than direct pointwise comparison of raw weather variables.

\subsection{Load-State Summary}
\label{subsec:load_state_summary}

The load-state branch converts a pure recent load segment into a compact numerical summary before similarity matching. This summary is intended to characterise the current operating state of the building, rather than to forecast the future load directly. For the query at forecast origin $t$, the input to the load-state encoder is the recent load-only sequence $\mathbf{y}^{\mathrm{past}}_t=[y_{t-L+1},\ldots,y_t]^\top$. For a historical memory item indexed by forecast origin $j$, the same construction gives $\mathbf{y}^{\mathrm{past}}_j=[y_{j-L+1},\ldots,y_j]^\top$ and $\mathbf{s}^{\mathrm{h}}_j=\phi_s(\mathbf{y}^{\mathrm{past}}_j)$.

In the default setting, $\phi_s(\cdot)$ maps the recent load segment to a 9-dimensional statistical state summary:
\begin{equation}
    \mathbf{s}^{\mathrm{q}}
    =
    \phi_s(\mathbf{y}^{\mathrm{past}}_t)
    =
    [
    y_t,\,
    \mu_t,\,
    \sigma_t,\,
    y^{\min}_t,\,
    y^{\max}_t,\,
    y_t-y_{t-L+1},\,
    \beta_t,\,
    \bar{\delta}^{(5)}_t,\,
    \bar{\delta}^{(11)}_t
    ]^\top
    \in \mathbb{R}^{9}.
\end{equation}
Here $\mu_t$, $\sigma_t$, $y^{\min}_t$, and $y^{\max}_t$ denote the mean, standard deviation, minimum, and maximum of $\mathbf{y}^{\mathrm{past}}_t$; $\beta_t$ denotes the linear trend slope fitted over the same segment; and $\bar{\delta}^{(m)}_t=m^{-1}\sum_{r=0}^{m-1}(y_{t-r}-y_{t-r-1})$ denotes the average recent first difference over the last $m$ steps, with $m\in\{5,11\}$. Thus, the summary summary combines the latest demand level, central tendency, dispersion, range, net change, long-window trend, and short-window ramping behaviour. After extraction, the summary is standardised using training-set statistics and $\ell_2$-normalised before cosine similarity is computed. In the pure-load-sequence ablation, $\phi_s(\cdot)$ is set to the identity mapping, and the normalised recent load window itself is used as the load-state summary.

\subsection{Retrieval Memory and Similarity Fusion}
\label{subsec:retrieval_database}

For each horizon $H$, WeL-RAC constructs a retrieval memory from training-set forecast origins whose full future horizon is observed. Since the horizon is fixed within one retrieval task, we write this memory simply as $\mathcal{M}$. Each memory item is indexed by a historical forecast origin $j$ and contains a historical load-state summary $\mathbf{s}^{\mathrm{h}}_j$ generated by the same encoder $\phi_s(\cdot)$, a weather-summary embedding generated as $\mathbf{z}^{\mathrm{h}}_j=g(\mathbf{W}^{\mathrm{h}}_j)$ from its stored horizon-aligned weather window, and the realised future load trajectory:
\begin{equation}
    \mathbf{r}_j
    =
    [y_{j+1}, \ldots, y_{j+H}]^\top .
\end{equation}
The memory is
\begin{equation}
    \mathcal{M}
    =
    \{(\mathbf{s}^{\mathrm{h}}_j,\mathbf{e}^{\mathrm{h}}_j,\mathbf{r}_j)\}_{j \in \mathcal{I}_{\mathrm{train}}},
\end{equation}
where $\mathbf{s}^{\mathrm{h}}_j$ is stored in the load-state bank, $\mathbf{e}^{\mathrm{h}}_j$ is stored in the load-aligned weather-summary bank, and $\mathbf{r}_j$ is the corresponding candidate analogue future. All weather-summary embeddings and load-state summarys are normalised before similarity computation as we mentioned before. The term load-aligned indicates that the weather summary and the realised future trajectory are associated with the same historical forecast origin and horizon.

\noindent Given the current query, the weather-summary similarity to memory item $j$ is computed as the cosine similarity:
\begin{equation}
    c^{\mathrm{w}}_{j}
    =
    (\mathbf{e}^{\mathrm{q}})^\top \mathbf{e}^{\mathrm{h}}_j.
\end{equation}
\noindent The load-state similarity is computed analogously:
\begin{equation}
    c^{\mathrm{s}}_{j}
    =
    (\mathbf{s}^{\mathrm{q}})^\top \mathbf{s}^{\mathrm{h}}_j .
\end{equation}
The two retrieval signals are then combined into a single analogue score:
\begin{equation}
    a_j
    =
    (1-\lambda)c^{\mathrm{w}}_{j}
    +
    \lambda c^{\mathrm{s}}_{j},
    \label{eq:retrieval_score}
\end{equation}
where $\lambda \in [0,1]$ controls the relative contribution of recent load-state matching. Setting $\lambda=0$ yields purely weather-summary retrieval, while $\lambda=1$ yields purely load-state retrieval. Intermediate values retrieve historical futures that are both meteorologically relevant and operationally compatible with the current building state.

\subsection{Retrieval of Analogue Futures}
\label{subsec:evidence_aggregation}

Let $\mathcal{N}_K$ be the set of top-$K$ memory items ranked by the analogue score in Eq.~\eqref{eq:retrieval_score}. Their contributions are normalised through a temperature-scaled softmax:
\begin{equation}
    \pi_j
    =
    \frac{
    \exp(a_j/\tau)
    }{
    \sum_{\ell \in \mathcal{N}_K}
    \exp(a_{\ell}/\tau)
    },
    \quad j \in \mathcal{N}_K,
\end{equation}
where $\tau>0$ controls the concentration of neighbour weights. The retrieved future evidence is the weighted average of the realised futures associated with the selected analogues:
\begin{equation}
    \widetilde{\mathbf{y}}
    =
    \sum_{j \in \mathcal{N}_K}
    \pi_j \mathbf{r}_j.
    \label{eq:retrieved_forecast}
\end{equation}
The resulting trajectory is a non-parametric analogue forecast induced from historical futures that resemble the current load state and weather regime.

\subsection{Post-prediction Retrieval-Augmented Correction}
\label{subsec:fusion}

The retrieved futures provides an evidence-based correction target for the initial prediction, as shown in Fig.~\ref{fig:method}. The final WeL-RAC prediction is
\begin{equation}
    \widehat{\mathbf{y}}
    =
    (1-\alpha)\widehat{\mathbf{y}}^{\mathrm{initial}}
    +
    \alpha \widetilde{\mathbf{y}}.
    \label{eq:final_fusion}
\end{equation}
Equation~\eqref{eq:final_fusion} instantiates the correction module $C(\cdot)$ as a lightweight weighted fusion. The correction coefficient $\alpha \in [0,1]$ controls the strength of retrieval correction for the current horizon. When $\alpha=0$, WeL-RAC reduces to the base forecaster; when $\alpha=1$, the final output relies entirely on retrieved analogue futures. Intermediate values provide conservative correction, which is useful when the base model is strong but can still benefit from historical evidence under similar weather-load conditions.

%% file: sections/03-data-preparation-hyperparameter-search.tex
\section{Experimental Setup} % check references
\label{sec:experimental_setup}
This section describes the datasets, forecasting protocol, baseline models, evaluation metric, and hyperparameter search spaces used to evaluate WeL-RAC. The goal is to ensure that the base forecasters and their retrieval-corrected variants are compared under the same chronological data split, forecasting horizons, and tuning procedure.

\begin{table}[H]
\centering
\caption{Overview of the datasets used in this study.}
\label{tab:dataset_summary}

\small
\renewcommand{\arraystretch}{1.18}

\begin{tabular*}{\linewidth}{
@{\extracolsep{\fill}}
c c c c c c c
@{}
}
\toprule
Dataset
& Type
& Scope
& Building ID
& Time span
& Resolution
& Records per building \\
\midrule

AlphaBuilding
& Synthetic
& 9 simulated buildings
& 1--9
& 2 years
& Hourly
& 17,520 \\

ASU
& Real-world
& 12 campus buildings
& 10--21
& 3 years
& Hourly
& 26,280 \\

UCF
& Real-world
& 10 campus buildings
& 22--31
& 3 years
& Hourly
& 26,278 \\

\bottomrule
\end{tabular*}
\end{table}

\subsection{Datasets and Preprocessing}
\label{subsec:datasets_preprocessing}
Table~\ref{tab:dataset_summary} summarizes the three benchmark datasets used in this study. We construct the benchmark datasets by selecting representative building cases from AlphaBuilding, Arizona State University (ASU), and University of Central Florida (UCF). For each case, the building load series are aligned with available building-related covariates and weather information, resulting in multivariate hourly time series that contain the forecasting target together with contextual information such as additional energy-use or operational variables and meteorological conditions. The exact data sources, selected buildings, and feature compositions differ across the three datasets and are described in Sections~\ref{Synthetic} and~\ref{Real-world}.

\subsubsection{Synthetic Dataset}
\label{Synthetic}

We use AlphaBuilding as the synthetic benchmark \citep{li2021synthetic}\footnote{AlphaBuilding dataset page: \url{https://lbnl-eta.github.io/AlphaBuilding-SyntheticDataset/}.}. The dataset contains simulated hourly building electricity consumption together with weather information under different climate and building-efficiency settings. In this study, we use nine two-year load--weather series from three cities representing distinct climate conditions: Miami with a hot and humid climate, San Francisco with a mild marine climate, and Chicago with cold winters and hot summers. For each city, three energy-efficiency configurations (High, Standard, and Low) are included, representing different levels of building energy performance. The corresponding periods, building IDs, and efficiency levels are summarized in Table~\ref{tab:alpha_subset}. Each building contains 17,520 hourly records, and the forecasting target OT is hourly building electricity consumption measured in kWh.

% Required packages:
% \usepackage{booktabs,tabularx,threeparttable}
\begin{table}[t]
\centering
\begin{threeparttable}
\caption{Composition of the AlphaBuilding dataset.}
\label{tab:alpha_subset}
\small
\setlength{\tabcolsep}{4pt}
\renewcommand{\arraystretch}{1.12}

\begin{tabular*}{\columnwidth}{
@{\extracolsep{\fill}} l l c c @{}
}
\toprule
City & Climate characteristics & Period & Building IDs (efficiency) \\
\midrule
Miami
& Hot and humid
& 2002--2003
& 1 (High) / 2 (Low) / 3 (Standard) \\

San Francisco
& Mild marine
& 2014--2015
& 4 (High) / 5 (Low) / 6 (Standard) \\

Chicago
& Cold winters and hot summers
& 2006--2007
& 7 (High) / 8 (Low) / 9 (Standard) \\
\bottomrule
\end{tabular*}

\end{threeparttable}
\end{table}

\subsubsection{Real-world Datasets}
\label{Real-world}
For the ASU benchmark, we select 12 campus buildings from the
Hierarchical Energy, Emissions, and Weather (HEEW) dataset
\citep{dong2025heew}\footnote{HEEW dataset:
\url{https://doi.org/10.6084/m9.figshare.28425647}.},
which integrates building-level energy records with weather
observations. The source dataset incorporates missing-value
imputation and anomaly correction as part of its original
data preparation. Our preprocessing restricts the records to
2017--2019, constructs hourly timestamps, and aligns the
building-level energy series with the shared weather observations.
The building files are then organized into a common multivariate
format, with electricity demand in kW designated as the prediction
target (\texttt{OT}) and PV generation, cooling and heating loads,
emissions, and observed weather variables retained as auxiliary
channels. Each resulting building series contains 26,280 hourly
records.

For the UCF real-world benchmark, we curate a subset from publicly available hourly utility-meter readings provided by the UCF Online Energy Information System (OEIS)\footnote{UCF OEIS building utility data: \url{https://oeis.ucf.edu/buildings}; UCF OEIS description: \url{https://energy.ucf.edu/2017/05/02/the-new-oeis-is-available/}.}. The raw time series data were not taken from BDG2; instead, BDG2 Panther/UCF records were used only as a metadata reference for building identifiers, use-type labels, floor area, construction year, site location, timezone, and climate-zone information \citep{miller2020building}\footnote{BDG2 repository: \url{https://github.com/buds-lab/building-data-genome-project-2}.}. After excluding buildings with substantial target missingness or unsuitable data quality, the processed UCF subset contains 10 campus buildings from 2017 to 2019 in US/Eastern local time. Orlando area weather observations were collected from the Iowa State University Iowa Environmental Mesonet archive\footnote{Iowa State IEM ASOS/METAR archive: \url{https://mesonet.agron.iastate.edu/request/download.phtml}.}. The UCF target OT is electricity demand converted to kWh, and the weather covariates include temperature, dew point, relative humidity, wind speed, wind gust, sea-level pressure, and precipitation. Each resulting building series contains 26,278 hourly
records. Missing wind gust values were set to zero, and remaining missing values were imputed by linear interpolation. %CHECK the data clean method

\subsubsection{Forecast Weather Inputs}
Because WeL-RAC constructs the weather-summary branch from horizon-aligned future weather information, each dataset is paired with a forecast-weather source. For AlphaBuilding, the future weather conditions associated with the simulated load trajectories are directly available. We therefore treat them as oracle weather forecasts, removing additional uncertainty from weather prediction and allowing the retrieval-correction mechanism to be evaluated more directly. This setting is used only for this synthetic dataset. For the ASU and UCF real-world benchmarks, external weather inputs are extracted from processed site-level High-Resolution Rapid Refresh (HRRR) forecast source. These sources cover 2017--2019 and provide forecast lead times from 1 to 18 hours at each site. HRRR is NOAA's hourly updated, convection-allowing, 3-km numerical weather prediction system for short-range forecasting \citep{dowell2022hrrr,james2022hrrr}. Notably, forecast origins without complete weather-forecast coverage under the adopted time-alignment policy are excluded from the evaluation for all methods, including the original forecaster, to ensure a consistent comparison set. At $H=18$, this excludes 4 forecast origins per ASU building (48 in total) and 18 per UCF building (180 in total).

\subsection{Baselines and Metrics}
\label{subsec:baselines_metrics}
We evaluate WeL-RAC as a model-agnostic correction layer attached to eight representative forecasting backbones. The selected backbones cover feed-forward neural networks, tree-based learning, recurrent networks, linear time series models, Transformer-style forecasters, MLP-based multiscale forecasters, exogenous-aware Transformer forecasters, and pretrained time series foundation models. Specifically, we use MLP \citep{rumelhart1986learning}, LightGBM \citep{ke2017lightgbm}, LSTM \citep{hochreiter1997long}, DLinear \citep{zeng2023transformers}, PatchTST \citep{nie2023patchtst}, TimeMixer \citep{wang2024timemixer}, TimeXer \citep{wang2024timexer}, and Chronos-2 \citep{ansari2025chronos2}. For each backbone, we compare the initial prediction with the corresponding retrieval-corrected prediction. This paired design directly measures whether the proposed retrieval-correction mechanism improves a given forecaster without changing its internal architecture.

Let $\{(y_i,\widehat{y}_i)\}_{i=1}^{N}$ denote all target--prediction pairs obtained by flattening the evaluated test forecast horizons for a given model, where $N$ is the total number of evaluated load values and $\bar{y}$ denotes the mean of the $N$ ground truth. We report four standard regression metrics: mean squared error (MSE), mean absolute error (MAE), coefficient of determination ($R^2$), and mean absolute percentage error (MAPE). They are defined as
\begin{equation}
    \mathrm{MSE}
    =
    \frac{1}{N}
    \sum_{i=1}^{N}
    (y_i-\widehat{y}_i)^2 ,
\end{equation}
\begin{equation}
    \mathrm{MAE}
    =
    \frac{1}{N}
    \sum_{i=1}^{N}
    |y_i-\widehat{y}_i| ,
\end{equation}
\begin{equation}
    R^2
    =
    1
    -
    \frac{\sum_{i=1}^{N}(y_i-\widehat{y}_i)^2}
    {\sum_{i=1}^{N}(y_i-\bar{y})^2} ,
\end{equation}
\begin{equation}
\mathrm{MAPE}
=
\frac{1}{|N|}
\sum_{i\in N}
\frac{|y_i-\hat{y}_i|}{|y_i|}.
\end{equation}
 Lower MSE, MAE, and MAPE indicate better performance, whereas higher $R^2$ indicates a better fit.

\begin{table}[H]
\centering
\caption{Hyperparameter search spaces for the eight forecasting models. Curly braces $\{\cdot\}$ denote discrete candidate values,
whereas square brackets $[\cdot]$ denote continuous search intervals.
For continuous \texttt{learning\_rate} intervals, sampling is performed on a logarithmic scale; discrete \texttt{learning\_rate} values are searched only over the listed candidates.}
\label{tab:hparam_search_spaces_all_models}
\scriptsize
\renewcommand{\arraystretch}{1.08}
\begin{tabularx}{\textwidth}{@{}p{0.13\textwidth}p{0.31\textwidth}X@{}}
\toprule
Model & Hyperparameter & Value range \\
\midrule
\multirow{9}{*}{MLP}
& \texttt{learning\_rate} & $[10^{-4}, 10^{-2}]$ \\
& \texttt{batch\_size} & $\{32, 64, 128\}$ \\
& \texttt{train\_epochs} & $\{10, 20, 30\}$ \\
& \texttt{hidden\_layers} & $\{1, 2, 3\}$ \\
& \texttt{hidden\_dims\_l1} when \texttt{hidden\_layers}=1 & $\{(16), (24), (32)\}$ \\
& \texttt{hidden\_dims\_l2} when \texttt{hidden\_layers}=2 & $\{(32,16), (48,24), (64,32)\}$ \\
& \texttt{hidden\_dims\_l3} when \texttt{hidden\_layers}=3 & $\{(48,32,16), (64,32,16), (64,64,32), (96,64,32)\}$ \\
& \texttt{dropout} & $\{0.0, 0.05, 0.10, 0.20\}$ \\
& \texttt{weight\_decay} & $\{0.0, 10^{-6}, 10^{-5}, 10^{-4}\}$ \\
\midrule
\multirow{9}{*}{LightGBM}
& \texttt{learning\_rate} & $[0.01, 0.10]$ \\
& \texttt{num\_leaves} & $\{15, 31, 63, 127\}$ \\
& \texttt{max\_depth} & $\{-1, 4, 6, 8, 10\}$ \\
& \texttt{min\_child\_samples} & $\{10, 20, 50, 100, 200\}$ \\
& \texttt{subsample} & $\{0.6, 0.8, 1.0\}$ \\
& \texttt{colsample\_bytree} & $\{0.6, 0.8, 1.0\}$ \\
& \texttt{reg\_lambda} & $\{0.0, 0.1, 1.0, 5.0, 10.0\}$ \\
& \texttt{reg\_alpha} & $\{0.0, 0.001, 0.01, 0.1, 1.0\}$ \\
& \texttt{min\_split\_gain} & $\{0.0, 0.01, 0.05, 0.1\}$ \\
\midrule
\multirow{8}{*}{PatchTST}
& \texttt{learning\_rate} & $[10^{-4}, 10^{-2}]$ \\
& \texttt{batch\_size} & $\{32, 64\}$ \\
& \texttt{train\_epochs} & $\{30\}$ \\
& \texttt{patch\_len} & $\{4, 6, 8\}$ \\
& \texttt{stride} & $\{2, 4\}$ \\
& \texttt{d\_model} & $\{64, 128\}$ \\
& \texttt{e\_layers} & $\{1, 2\}$ \\
& \texttt{dropout} & $\{0.1, 0.2\}$ \\
& \texttt{fc\_dropout} & $\{0.1, 0.2\}$ \\
\midrule
\multirow{4}{*}{DLinear}
& \texttt{learning\_rate} & $[10^{-4}, 10^{-2}]$ \\
& \texttt{batch\_size} & $\{32, 64, 128\}$ \\
& \texttt{train\_epochs} & $\{30\}$ \\
& \texttt{moving\_avg} & $\{3, 5, 7, 13\}$ \\
& \texttt{individual} & $\{\texttt{False}, \texttt{True}\}$ \\
\midrule
\multirow{5}{*}{LSTM}
& \texttt{learning\_rate} & $[10^{-4}, 10^{-2}]$ \\
& \texttt{batch\_size} & $\{32, 64, 128\}$ \\
& \texttt{train\_epochs} & $\{30\}$ \\
& \texttt{hidden\_size} & $\{32, 64, 128\}$ \\
& \texttt{num\_layers} & $\{1, 2\}$ \\
& \texttt{dropout} & $\{0.0, 0.1, 0.2\}$ \\
\midrule
\multirow{9}{*}{TimeMixer}
& \texttt{learning\_rate} & $[10^{-4}, 10^{-2}]$ \\
& \texttt{batch\_size} & $\{32, 64\}$ \\
& \texttt{train\_epochs} & $\{30\}$ \\
& \texttt{d\_model} & $\{64, 128\}$ \\
& \texttt{e\_layers} & $\{1, 2\}$ \\
& \texttt{down\_sampling\_pair} $(\texttt{layers}, \texttt{window})$ & $\{(1,2), (1,3), (2,2)\}$ \\
% & \texttt{down\_sampling\_layers} derived from \texttt{down\_sampling\_pair} & $\{1, 2\}$ \\
% & \texttt{down\_sampling\_window} derived from \texttt{down\_sampling\_pair} & $\{2, 3\}$ \\
& \texttt{moving\_avg} & $\{3, 5, 7\}$ \\
& \texttt{dropout} & $\{0.0, 0.1, 0.2\}$ \\
\midrule
\multirow{4}{*}{TimeXer}
& \texttt{learning\_rate} & $\{10^{-4}, 10^{-3}, 10^{-2}\}$ \\
& \texttt{batch\_size} & $\{32, 64\}$ \\
& \texttt{train\_epochs} & $\{30\}$ \\
& \texttt{e\_layers} & $\{1, 2, 3\}$ \\
& \texttt{d\_model} & $\{128, 256, 512\}$ \\
\bottomrule
\end{tabularx}
\end{table}

\begin{table}[H]
\centering
\caption{Hyperparameter search space for the retrieval branch and correction module.}
\label{tab:hparam_search_space_retrieval_branch}
\footnotesize
\renewcommand{\arraystretch}{1.08}
\begin{tabularx}{\linewidth}{@{}p{0.28\linewidth}p{0.32\linewidth}X@{}}
\toprule
Component & Hyperparameter & Value range \\
\midrule
\multirow{3}{*}{Retrieval Branch}
& $\lambda$ (\texttt{state weight}) & $\{0.20, 0.35, 0.50, 0.65, 0.80\}$ \\
& $K$ (\texttt{top\_K}) & $\{10, 15, 20, 30\}$ \\
& $\tau$ (\texttt{temperature}) & $\{0.10, 0.20, 0.30\}$ \\
\midrule
Correction Module
& $\alpha$ (\texttt{correction coefficient}) & $\{0.0, 0.1, 0.2, 0.3, 0.4, 0.5, 0.6, 0.7, 0.8, 0.9, 1.0\}$ \\
\bottomrule
\end{tabularx}
\end{table}

\subsection{Forecasting Protocol} %LightGBM used future forecasting to output prediction
\label{subsec:forecasting_protocol}
The forecasting protocol defines how each prediction task is constructed and evaluated. For each building and each prediction horizon $H \in {1, 6, 12, 18}$, the time series is split chronologically into training, validation, and test sets with a ratio of 7/1/2. For all trainable forecasting backbones, the training period is used to fit the forecasting model and construct the retrieval memory, while the validation period is used to select both forecasting-model hyperparameters and retrieval-correction hyperparameters. Chronos-2 is evaluated in its pretrained zero-shot configuration and is therefore neither trained nor fine-tuned on the building datasets; only its associated retrieval-correction hyperparameters are selected on the validation period. The test period is held out until the final evaluation.

At each forecast origin $t$, the base forecaster receives the historical load-covariate window $\mathbf{X}$ and produces the initial prediction $\widehat{\mathbf{y}}^{\mathrm{initial}}$. The retrieval branch uses the same forecast origin to construct the load-state query $\mathbf{s}^{\mathrm{q}}$ from the recent load segment and the weather-summary query $\mathbf{z}^{\mathrm{q}}$ from the horizon-aligned weather window. Retrieval is performed only against training-period memory items, and the retrieved analogue future is fused with the initial prediction using the correction rule in Eq.~\eqref{eq:final_fusion}. This protocol prevents validation or test future load trajectories from entering the retrieval memory.

To ensure consistent paired comparisons, each base forecaster and
its retrieval-corrected part use the same chronological data
split, input length, prediction horizon, random seed settings, and
evaluation windows. The main experiments use a common 24-hour input
window for all forecasting backbones. 
% Additional analyses of
% Chronos-2's context length and historical organization are provided
% in the Supplementary Material.
For LightGBM,
horizon-aligned weather forecasts are additionally included as
future exogenous inputs. All reported results are obtained on the test set after selecting the forecaster and retrieval-correction hyperparameters on the validation set.

\subsection{Hyperparameter Search}
\label{subsec:hyperparameter_search}
Hyperparameter selection is performed only on the validation period defined in Sec.~\ref{subsec:forecasting_protocol}. We separate the search into two parts. First, each trainable forecasting backbone is tuned using the candidate spaces in Table~\ref{tab:hparam_search_spaces_all_models}; Chronos-2 is evaluated with its fixed pretrained inference configuration and is therefore not included in this search table. Second, after the initial forecasts are obtained, the retrieval-correction branch is tuned using the candidate spaces in Table~\ref{tab:hparam_search_space_retrieval_branch}. The best configuration is selected according to validation performance under the primary metric, and the held-out test period is used only for final reporting. This two-stage procedure keeps the base forecast generation and retrieval correction comparable across paired initial and corrected predictions.

%% file: sections/04-experiments-results.tex
\section{Experimental Results}
\label{sec:results}

This section evaluates WeL-RAC on the synthetic AlphaBuilding
dataset and the two real-world campus datasets, ASU and UCF.
We first compare the initial and retrieval-corrected predictions
of eight forecasting backbones at the dataset level, followed
by analyses across individual buildings and forecasting horizons.
We then conduct ablation studies and sensitivity analysis to
examine the contributions of individual components and the
effects of key retrieval and fusion parameters. Finally, we
assess the computational cost of the retrieval-correction module.

\subsection{Overall Forecasting Performance}
\label{subsec:overall_forecasting_performance}

As shown in Table~\ref{tab:dataset_level_results}, the proposed retrieval mechanism yields highly consistent improvements across the three building datasets and eight forecasting backbones. Among the original forecasters, LightGBM provides the strongest overall initial predictions on ASU and UCF, achieving the lowest MSE and MAE on both datasets, while PatchTST achieves the lowest initial MSE on AlphaBuilding. After incorporating retrieval, the enhanced models outperform their corresponding base forecasters in 90 of the 96 comparisons across datasets and metrics. In particular, MSE, MAE, and \(R^2\) improve for every combination of forecasting backbone and dataset, demonstrating that the proposed mechanism is broadly effective across heterogeneous forecasting architectures rather than being tailored to a specific model class. The retrieval mechanism also changes the relative ranking of the forecasting backbones: LightGBM+Retrieval remains the strongest model on ASU in terms of MSE and MAE, whereas TimeMixer+Retrieval achieves the lowest MSE on both AlphaBuilding and UCF, with values of 74.674 and 246.582, respectively; LightGBM+Retrieval achieves the lowest MAE across all three datasets. The gains are particularly pronounced on AlphaBuilding, where models with comparatively weak initial predictions benefit substantially; for example, given the poor accuracy of the initial forecasts, the selected correction weight is $\alpha = 1$ for both DLinear and Chronos-2, effectively reducing the framework to retrieval-only forecasting. Their MSEs decrease from 540.429 to 112.013 and from 703.045 to 112.013, respectively. Improvements on ASU and UCF are generally more moderate but remain highly consistent for MSE, MAE, and \(R^2\). MAPE is the main exception, particularly on UCF, where retrieval occasionally increases the relative percentage error despite improving the other three metrics. Overall, these results indicate that retrieval provides complementary, instance-specific historical information that can robustly improve building load forecasts without modifying the underlying forecasting architecture.

\begin{table*}[t]
\centering
\caption{Overall forecasting performance. For each building and horizon, all forecasters and their retrieval-enhanced counterparts are evaluated on exactly the same retained forecast origins. Metrics are computed separately for each building--horizon--seed combination and then equally averaged over all buildings, four forecasting horizons ($H\in\{1,6,12,18\}$), and three random seeds ($2021$, $0$, and $42$) within each dataset. All models use an input length of 24. Lower values are better for MSE, MAE, and MAPE, whereas higher values are better for $R^2$. For each backbone, the better result between the original forecaster and its retrieval-enhanced counterpart is highlighted in \textbf{bold}. The Best column reports the number of better results among the 12 dataset--metric comparisons for each forecaster and retrieval pair. All displayed values are rounded to three decimal places; boldface and Best counts are determined using the original unrounded values.}
\label{tab:dataset_level_results}
\begin{threeparttable}
\setlength{\tabcolsep}{3.0pt}
\renewcommand{\arraystretch}{1.08}
\resizebox{\textwidth}{!}{%
\begin{tabular}{l|cccc|cccc|cccc|c}
\toprule
\multirow{2}{*}{Method}
& \multicolumn{4}{c|}{ASU}
& \multicolumn{4}{c|}{AlphaBuilding}
& \multicolumn{4}{c|}{UCF}
& \multirow{2}{*}{Best} \\
\cmidrule(lr){2-5}
\cmidrule(lr){6-9}
\cmidrule(lr){10-13}
& MSE$\downarrow$ & MAE$\downarrow$ & $R^2\uparrow$ & MAPE$\downarrow$
& MSE$\downarrow$ & MAE$\downarrow$ & $R^2\uparrow$ & MAPE$\downarrow$
& MSE$\downarrow$ & MAE$\downarrow$ & $R^2\uparrow$ & MAPE$\downarrow$
& \\
\midrule
MLP
& 543.840 & 12.382 & 0.753 & 0.064
& 105.593 & 5.205 & 0.947 & 0.093
& 291.787 & 9.117 & 0.789 & \textbf{0.058}
& 1 \\
MLP+Retrieval
& \textbf{532.139} & \textbf{12.237} & \textbf{0.773} & \textbf{0.063}
& \textbf{93.421} & \textbf{4.577} & \textbf{0.952} & \textbf{0.080}
& \textbf{280.019} & \textbf{9.094} & \textbf{0.793} & 0.064
& \textbf{11} \\
\midrule
LightGBM
& 480.667 & 11.002 & 0.784 & 0.057
& 129.410 & 4.413 & 0.932 & 0.073
& 256.371 & 8.620 & 0.799 & 0.073
& 0 \\
LightGBM+Retrieval
& \textbf{474.789} & \textbf{10.974} & \textbf{0.792} & \textbf{0.057}
& \textbf{81.687} & \textbf{4.181} & \textbf{0.957} & \textbf{0.071}
& \textbf{250.241} & \textbf{8.544} & \textbf{0.802} & \textbf{0.073}
& \textbf{12} \\
\midrule
DLinear
& 630.605 & 13.476 & 0.777 & 0.067
& 540.429 & 13.798 & 0.728 & 0.236
& 351.967 & 9.957 & 0.743 & \textbf{0.063}
& 1 \\
DLinear+Retrieval
& \textbf{561.233} & \textbf{12.953} & \textbf{0.799} & \textbf{0.065}
& \textbf{112.013} & \textbf{4.814} & \textbf{0.942} & \textbf{0.081}
& \textbf{309.098} & \textbf{9.781} & \textbf{0.764} & 0.070
& \textbf{11} \\
\midrule
LSTM
& 658.946 & 14.743 & 0.724 & 0.077
& 85.760 & 4.474 & 0.957 & 0.079
& 350.110 & 10.631 & 0.718 & 0.079
& 0 \\
LSTM+Retrieval
& \textbf{594.805} & \textbf{13.618} & \textbf{0.759} & \textbf{0.071}
& \textbf{77.218} & \textbf{4.220} & \textbf{0.961} & \textbf{0.074}
& \textbf{313.508} & \textbf{10.086} & \textbf{0.743} & \textbf{0.077}
& \textbf{12} \\
\midrule
PatchTST
& 589.027 & 13.917 & 0.763 & 0.072
& 84.651 & 4.831 & 0.957 & 0.086
& 343.086 & 10.585 & 0.731 & 0.081
& 0 \\
PatchTST+Retrieval
& \textbf{558.501} & \textbf{13.185} & \textbf{0.778} & \textbf{0.068}
& \textbf{77.219} & \textbf{4.416} & \textbf{0.961} & \textbf{0.077}
& \textbf{311.510} & \textbf{10.058} & \textbf{0.750} & \textbf{0.079}
& \textbf{12} \\
\midrule
TimeMixer
& 544.753 & 11.828 & 0.789 & 0.060
& 100.152 & 5.252 & 0.950 & 0.090
& 259.768 & 8.684 & 0.792 & \textbf{0.057}
& 1 \\
TimeMixer+Retrieval
& \textbf{501.236} & \textbf{11.512} & \textbf{0.809} & \textbf{0.059}
& \textbf{74.674} & \textbf{4.526} & \textbf{0.962} & \textbf{0.077}
& \textbf{246.582} & \textbf{8.586} & \textbf{0.802} & 0.061
& \textbf{11} \\
\midrule
TimeXer
& 554.301 & 11.862 & 0.806 & 0.060
& 123.162 & 6.193 & 0.937 & 0.114
& 283.434 & 9.068 & 0.789 & \textbf{0.057}
& 1 \\
TimeXer+Retrieval
& \textbf{527.180} & \textbf{11.682} & \textbf{0.815} & \textbf{0.060}
& \textbf{96.709} & \textbf{4.874} & \textbf{0.950} & \textbf{0.087}
& \textbf{266.839} & \textbf{8.957} & \textbf{0.798} & 0.062
& \textbf{11} \\
\midrule
Chronos-2
& 740.949 & 14.012 & 0.741 & \textbf{0.067}
& 703.045 & 14.558 & 0.648 & 0.231
& 383.182 & 9.990 & 0.733 & \textbf{0.061}
& 2 \\
Chronos-2+Retrieval
& \textbf{568.597} & \textbf{13.362} & \textbf{0.787} & 0.067
& \textbf{112.013} & \textbf{4.814} & \textbf{0.942} & \textbf{0.081}
& \textbf{320.879} & \textbf{9.910} & \textbf{0.764} & 0.071
& \textbf{10} \\
\bottomrule
\end{tabular}%
}
% \begin{tablenotes}[flushleft]
% \footnotesize
% \item \textit{Evaluation subset:}
% Windows without complete forecast-weather coverage under the current
% time-alignment policy are excluded from every method, including the
% original forecasters. At $H=18$, this excludes 4 windows per ASU
% building (48 in total) and 18 per UCF building (180 in total);
% no AlphaBuilding windows or windows at the other horizons are excluded.
% The UCF exclusions arise from daylight-saving-time alignment and the
% 18-hour forecast-coverage limit, rather than missing numerical
% weather values.
% \item \textit{Reporting:}
% Chronos-2 uses Recent24 inputs. MAPE is reported as a ratio,
% not a percentage. Identical displayed values can have different
% boldface because comparisons use unrounded values.
% \end{tablenotes}
\end{threeparttable}
\end{table*}

% building 
\subsection{Building-wise Results}

\begin{figure}
    \centering
    \includegraphics[width=0.98\linewidth]
    {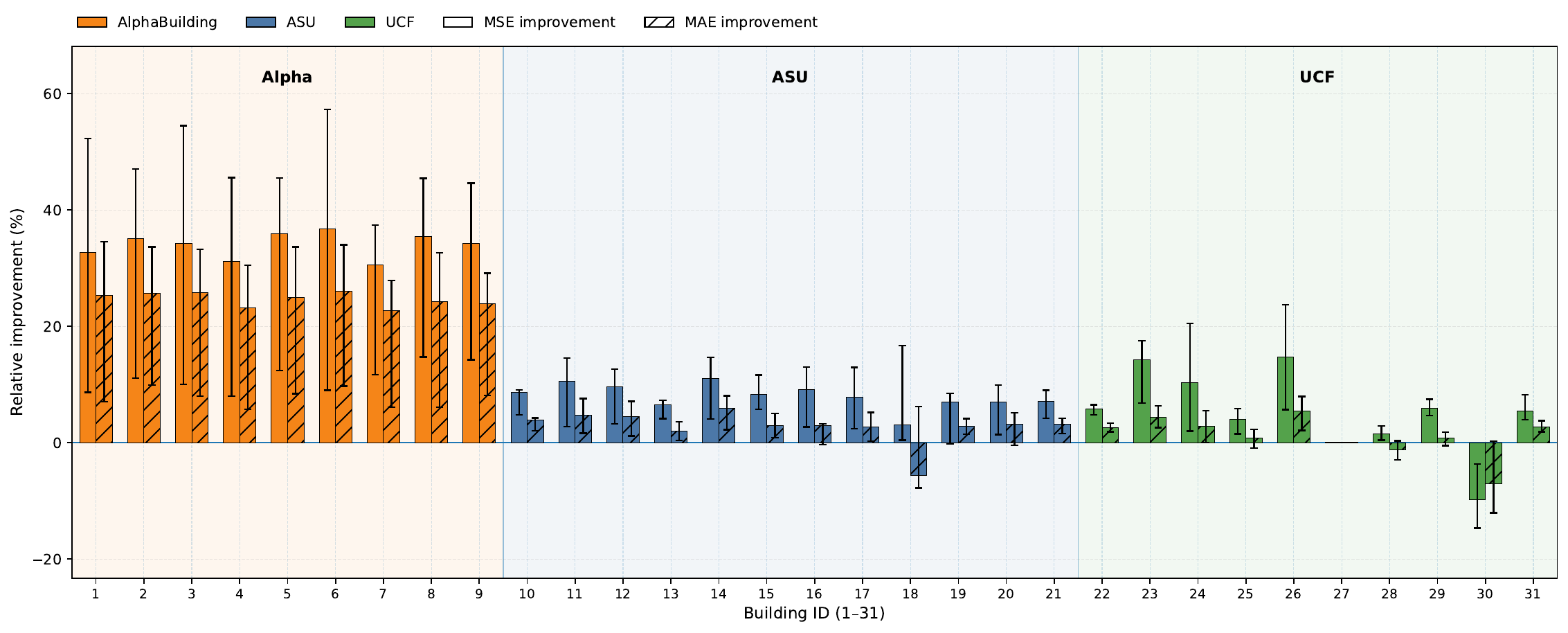}
    \caption{
    Building-wise relative improvements of the retrieval-enhanced forecasts over the corresponding original forecasters.
    Results are averaged over the four forecasting horizons
    ($H\in\{1,6,12,18\}$) and all eight forecasting backbones.
    Solid bars denote the relative improvement in MSE, while hatched bars denote the relative improvement in MAE.
    Positive values indicate that retrieval improves the original forecast.
    Building IDs 1--9, 10--21, and 22--31 correspond to AlphaBuilding, ASU, and UCF, respectively.
    }
    \label{fig:building_wise_improvement}
\end{figure}

Figure~\ref{fig:building_wise_improvement} evaluates retrieval gains at the building level, with relative MSE and MAE improvements averaged over all four forecasting horizons and eight forecasting backbones. Positive values indicate that retrieval improves the original forecast. Overall, retrieval improves MSE for 30 of the 31 buildings and MAE for 28 of the 31 buildings, indicating that the gains are broadly distributed across buildings rather than being driven by a few isolated cases.

The improvements are most consistent on AlphaBuilding (Buildings~1--9), where all buildings benefit, with average reductions of approximately $34.0\%$ in MSE and $24.6\%$ in MAE. ASU (Buildings~10--21) also shows stable gains, averaging about $8.0\%$ in MSE and $2.8\%$ in MAE, except building 18. UCF (Buildings~22--31) is more heterogeneous, with average improvements of approximately $5.2\%$ in MSE and $1.1\%$ in MAE, except building 29 and 30. Only a small number of UCF buildings exhibit degradation, with Building~30 being the clearest failure case. Overall, these results demonstrate that the proposed retrieval mechanism provides robust building-wise improvements across heterogeneous datasets. The larger gains on AlphaBuilding may be attributed to its synthetic nature, where more regular and repeatable patterns make similarity-based retrieval more likely to identify informative historical cases. In contrast, the real-world ASU and UCF datasets exhibit more complex and variable dynamics, such that historically similar conditions do not necessarily imply similar future load trajectories. We further investigate the factors underlying the success and failure of retrieval in Section~\ref{sec:case_study}. % add section link?

\subsection{Horizon-wise Results}
\begin{figure}
    \centering
    \includegraphics[width=\textwidth]{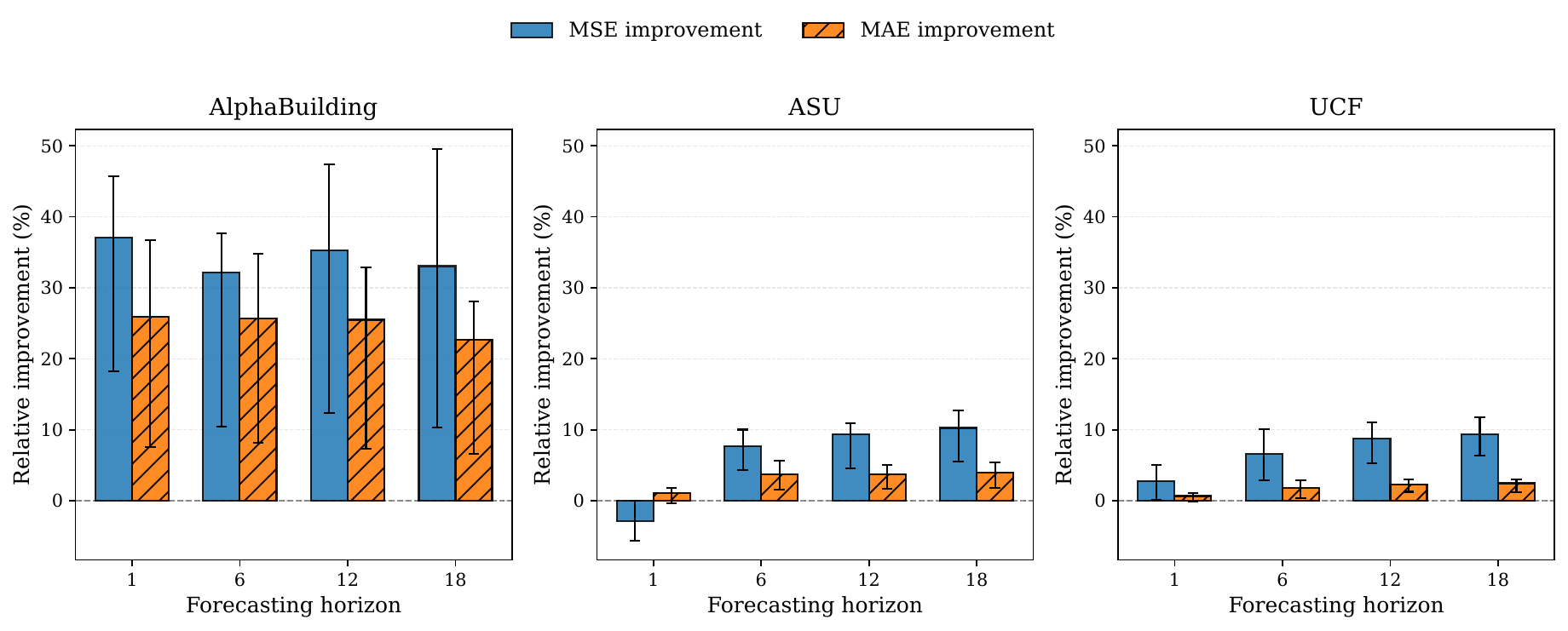}
    \caption{
    Horizon-wise relative improvements of the retrieval-enhanced forecasts over the original forecasters for MSE and MAE.
    Results are averaged over all buildings within each dataset and over all eight forecasting backbones.
    In each panel, each forecasting horizon contains two bars: the plain bar denotes the MSE improvement and the hatched bar denotes the MAE improvement.
    Error bars indicate the interquartile range across forecasting backbones.
    Positive values indicate that retrieval improves the original forecast.
    }
    \label{fig:horizon_mse_mae_bar}
\end{figure}

Figure~\ref{fig:horizon_mse_mae_bar} presents the horizon-wise relative improvements brought by retrieval, averaged over all buildings and all eight forecasting backbones.
AlphaBuilding shows consistently large gains across all forecasting horizons, with MSE improvements remaining above $32\%$ and MAE improvements above $22\%$.
This indicates that retrieval provides strong and stable benefits for the synthetic benchmark over both short- and longer-horizon forecasting.

For the two real-world datasets, the effect of retrieval becomes more pronounced as the forecasting horizon increases.
On ASU, the average MSE improvement changes from $-2.9\%$ at $H=1$ to $7.7\%$, $9.3\%$, and $10.3\%$ at $H=6,12,18$, respectively, while the MAE improvement increases from $1.1\%$ to around $3.7$--$3.9\%$.
UCF exhibits a similar trend, with MSE improvements growing from $2.8\%$ at $H=1$ to $6.7\%$, $8.8\%$, and $9.4\%$ at longer horizons, and MAE improvements rising from $0.7\%$ to $2.5\%$.
Overall, these results suggest that retrieval is particularly beneficial in longer-horizon forecasting, where the base forecasters face greater uncertainty and historical analogues provide more useful corrective information.

% Required packages in the manuscript preamble:
% \usepackage{booktabs,multirow,graphicx,threeparttable}

% Required packages in the manuscript preamble:
% \usepackage{booktabs,multirow,graphicx,threeparttable}

\subsection{Comparison with Existing Retrieval-Based Methods}
\label{sec:retrieval_baseline_comparison}

We compare WeL-RAC with RAFT~\cite{han2025raft}, SERAF~\cite{zhou2026seraf}, and SARAF~\cite{zhou2026stationarity}. All three retrieval-based baseline
implementations are MLP-based. And we searched hyperparameters on the
validation set for each building--horizon task, rather than default
settings. We therefore choose WeL-RAC with an MLP backbone to compare
methods within the same forecaster family. Their input information is,
however, not identical: the full WeL-RAC uses query-aligned future weather
in its semantic-weather retrieval branch, whereas the three baselines
use historical inputs without access to query-aligned future weather.
To control for this difference in retrieval information, we additionally
evaluate a load-only version of Wel-RAC, which removes the
semantic-weather branch and retrieves using only historical load-state
summaries. This provides a comparison without the additional
future-weather information supplied through the retrieval branch.

\begin{table*}[t]
\centering
\caption{Comparison with retrieval baselines. Load only denotes retrieval based solely on the proposed load-state summary of WeL-RAC, whereas the full version uses both retrieval branches. For each building and horizon, all methods are evaluated on the same retained forecast origins. Metrics are computed separately for each building--horizon--seed combination and then averaged equally over all buildings, four forecasting horizons ($H\in\{1,6,12,18\}$), and three random seeds within each dataset. All models use an input length of 24. Lower values are better for MSE, MAE, and MAPE, whereas higher values are better for $R^2$. The best result among the five rows in each dataset--metric column is highlighted in \textbf{bold}. The Best column counts column-wise wins; both boldface and counts are determined using unrounded values.}
\label{tab:retrieval_baseline_comparison}
\begin{threeparttable}
\setlength{\tabcolsep}{3.0pt}
\renewcommand{\arraystretch}{1.08}
\resizebox{\textwidth}{!}{%
\begin{tabular}{l|cccc|cccc|cccc|c}
\toprule
\multirow{2}{*}{Method}
& \multicolumn{4}{c|}{ASU}
& \multicolumn{4}{c|}{AlphaBuilding}
& \multicolumn{4}{c|}{UCF}
& \multirow{2}{*}{Best} \\
\cmidrule(lr){2-5}
\cmidrule(lr){6-9}
\cmidrule(lr){10-13}
& MSE$\downarrow$ & MAE$\downarrow$ & $R^2\uparrow$ & MAPE$\downarrow$
& MSE$\downarrow$ & MAE$\downarrow$ & $R^2\uparrow$ & MAPE$\downarrow$
& MSE$\downarrow$ & MAE$\downarrow$ & $R^2\uparrow$ & MAPE$\downarrow$
& \\
\midrule
RAFT
& 577.894 & 13.215 & 0.785 & 0.068
& 299.024 & 10.925 & 0.850 & 0.202
& 317.783 & 9.794 & 0.759 & 0.062
& 0 \\
SERAF
& 543.535 & 12.718 & \textbf{0.798} & 0.065
& 249.901 & 9.864 & 0.874 & 0.178
& 295.230 & 9.467 & 0.775 & \textbf{0.061}
& 2 \\
SARAF
& 546.905 & 12.857 & 0.795 & 0.066
& 439.627 & 13.630 & 0.781 & 0.254
& 291.016 & 9.452 & 0.782 & 0.061
& 0 \\
\midrule
MLP+Retrieval (load only)
& 535.118 & 12.352 & 0.767 & 0.064
& 100.142 & 4.771 & 0.949 & 0.082
& 288.121 & 9.202 & 0.787 & 0.063
& 0 \\
MLP+Retrieval (full)
& \textbf{532.139} & \textbf{12.237} & 0.773 & \textbf{0.063}
& \textbf{93.421} & \textbf{4.577} & \textbf{0.952} & \textbf{0.080}
& \textbf{280.019} & \textbf{9.094} & \textbf{0.793} & 0.064
& \textbf{10} \\
\bottomrule
\end{tabular}%
}
\end{threeparttable}
\end{table*}

Table~\ref{tab:retrieval_baseline_comparison} shows that the load-only
variant achieves lower MSE and MAE than all three baselines on every
dataset. Relative to the strongest baseline for each metric, its MSE
reductions are 1.55\%, 59.93\%, and 0.99\%, and its MAE reductions are
2.88\%, 51.63\%, and 2.64\% on ASU, AlphaBuilding, and UCF, respectively.
Thus, competitive MSE and MAE are retained without semantic-weather
retrieval. Enabling the weather branch further reduces MSE by 0.56\%,
6.71\%, and 2.81\% relative to the load-only variant on the three
datasets, respectively. The full model achieves the best results in
10 of the 12 dataset--metric comparisons; SERAF only remains best for
$R^2$ on ASU and MAPE on UCF. These results suggest that WeL-RAC is better suited to building load consumption forecasting than the evaluated general retrieval-based
methods. Its load-only variant already achieves lower MSE and MAE across
all three datasets, while enabling the semantic-weather branch further
improves both metrics, highlighting the value of incorporating
domain-specific information into retrieval.

\subsection{Ablation Studies}

\begin{table*}[t]
\centering
\caption{Dataset-level ablation results averaged over eight forecasting models.
For each model, results are first averaged over all buildings,
four forecasting horizons ($H\in\{1,6,12,18\}$), and three random seeds;
the reported values are then macro-averaged across the eight models. Lower values are better for MSE, MAE, and MAPE, whereas higher values
are better for $R^2$. Boldface denotes the best result and underlining denotes the
second-best result within each column.}
\label{tab:ablation_all8_dataset}

\small
\setlength{\tabcolsep}{3.5pt}
\renewcommand{\arraystretch}{1.08}

\resizebox{\textwidth}{!}{%
\begin{tabular}{lccccccccccccr}
\toprule
\multirow[c]{2}{*}{Method}
& \multicolumn{4}{c}{ASU}
& \multicolumn{4}{c}{AlphaBuilding}
& \multicolumn{4}{c}{UCF}
& \multirow[c]{2}{*}{Best} \\
\cmidrule(lr){2-5}
\cmidrule(lr){6-9}
\cmidrule(lr){10-13}

& MSE$\downarrow$ & MAE$\downarrow$ & $R^2\uparrow$ & MAPE$\downarrow$
& MSE$\downarrow$ & MAE$\downarrow$ & $R^2\uparrow$ & MAPE$\downarrow$
& MSE$\downarrow$ & MAE$\downarrow$ & $R^2\uparrow$ & MAPE$\downarrow$
& \\
\midrule

Full
& \textbf{539.810} & \textbf{12.440} & \textbf{0.789} & \textbf{0.064}
& 90.619 & 4.553 & 0.954 & 0.079
& \textbf{287.335} & \textbf{9.377} & \textbf{0.777} & 0.070
& \textbf{7} \\

Forecaster only
& 592.886 & 12.903 & 0.767 & 0.065
& 234.025 & 7.340 & 0.882 & 0.125
& 314.963 & 9.582 & 0.762 & \textbf{0.066}
& 1 \\

Retrieval only
& 674.790 & 14.834 & 0.729 & 0.075
& 112.735 & 4.830 & 0.942 & 0.081
& 394.999 & 11.310 & 0.681 & 0.091
& 0 \\

Load summary only
& \underline{542.388} & \underline{12.510}
& \underline{0.787} & \underline{0.064}
& 96.977 & 4.732 & 0.951 & 0.081
& \underline{293.568} & \underline{9.451}
& \underline{0.774} & \underline{0.069}
& 0 \\

Load raw sequence only
& 554.645 & 12.835 & 0.780 & 0.065
& \underline{83.288} & \underline{4.318}
& \underline{0.958} & \textbf{0.072}
& 325.249 & 9.941 & 0.764 & 0.071
& 1 \\

Weather summary + load raw
& 553.016 & 12.823 & 0.780 & 0.065
& \textbf{82.853} & \textbf{4.269}
& \textbf{0.958} & \underline{0.073}
& 321.613 & 9.906 & 0.765 & 0.073
& \textbf{3} \\

Weather numeric + load summary
& 606.374 & 13.744 & 0.762 & 0.070
& 118.535 & 5.425 & 0.939 & 0.095
& 372.322 & 10.768 & 0.734 & 0.079
& 0 \\

Weather full summary only
& 919.753 & 19.244 & 0.639 & 0.093
& 610.668 & 15.662 & 0.694 & 0.277
& 440.846 & 12.183 & 0.671 & 0.092
& 0 \\

Weather numeric only
& 1053.586 & 20.879 & 0.586 & 0.110
& 658.023 & 17.283 & 0.671 & 0.304
& 540.277 & 13.638 & 0.601 & 0.097
& 0 \\

\bottomrule
\end{tabular}%
}
\end{table*}

\noindent
The ablation variants are defined as follows.
\textit{Full} is our model Wel-RAC.
\textit{Forecaster only} removes retrieval correction entirely
($\alpha=0$), while \textit{Retrieval only} removes the forecaster
contribution and relies entirely on retrieval ($\alpha=1$).
\textit{Load summary only} performs retrieval using only the proposed
load-state summary, whereas \textit{Load raw sequence only} uses the
raw historical load sequence as the sole retrieval representation.
\textit{Weather summary + load raw} retains the proposed weather
summary but replaces the load-state summary with the raw load sequence.
Conversely, \textit{Weather numeric + load summary} retains the
load-state summary but replaces the proposed weather summary with
direct numeric weather features.
Finally, \textit{Weather summary only} and
\textit{Weather numeric only} remove the load-state branch and perform
retrieval using only the corresponding weather representation.
Unless explicitly changed by an ablation, the selected retrieval and
correction parameters are kept unchanged.

\paragraph{Overall Ablation analysis.}
Table~\ref{tab:ablation_all8_dataset} shows that the \textit{Full}
provides the most robust overall performance across the eight
forecasting models, achieving the best result in 7 of the 12
across dataset and metric comparisons.
In particular, it is best across all four metrics on ASU and achieves
the lowest MSE and MAE and the highest $R^2$ on UCF.
\textit{Load summary only} is consistently the strongest competing variant on
the two real-world datasets, whereas both weather-only variants are
substantially weaker.
Moreover, replacing the proposed weather summary with direct numeric
weather inputs while keeping the load summary unchanged consistently
degrades performance on all three datasets.
Together, these results indicate that load-state similarity provides
the primary retrieval signal, while the structured weather
representation supplies complementary information that improves the
reliability of the correction.

\paragraph{Effect of load representation.}
The preferred load representation, however, depends on the dataset.
On ASU and UCF, summary-based state matching is more effective than
matching the complete historical load trajectory.
In these real-world datasets, a similar load shape does not necessarily
correspond to the same operating state or lead to a similar future load,
whereas the state summary provides a more compact representation of the
operating condition.
AlphaBuilding exhibits the opposite behavior: its historical load
profiles show stronger shape repetition, so retaining the complete load
sequence makes it easier to identify reusable recurring patterns.
Consequently, replacing the load summary with the raw sequence while
keeping the weather representation unchanged yields the best MSE, MAE,
and $R^2$ on AlphaBuilding.
This contrast suggests that load summaries are beneficial when
shape similarity is an imperfect proxy for operating state, whereas
raw sequence matching can be preferable when historical load patterns
are highly repetitive. In addition, the proposed load-state representation compresses each
retrieval item from a 24-dimensional raw load sequence to a
nine-dimensional state vector, reducing memory usage while preserving the
information needed for effective state matching.

\subsection{Sensitivity Analysis}

\begin{figure}
    \centering
    \includegraphics[width=\textwidth]{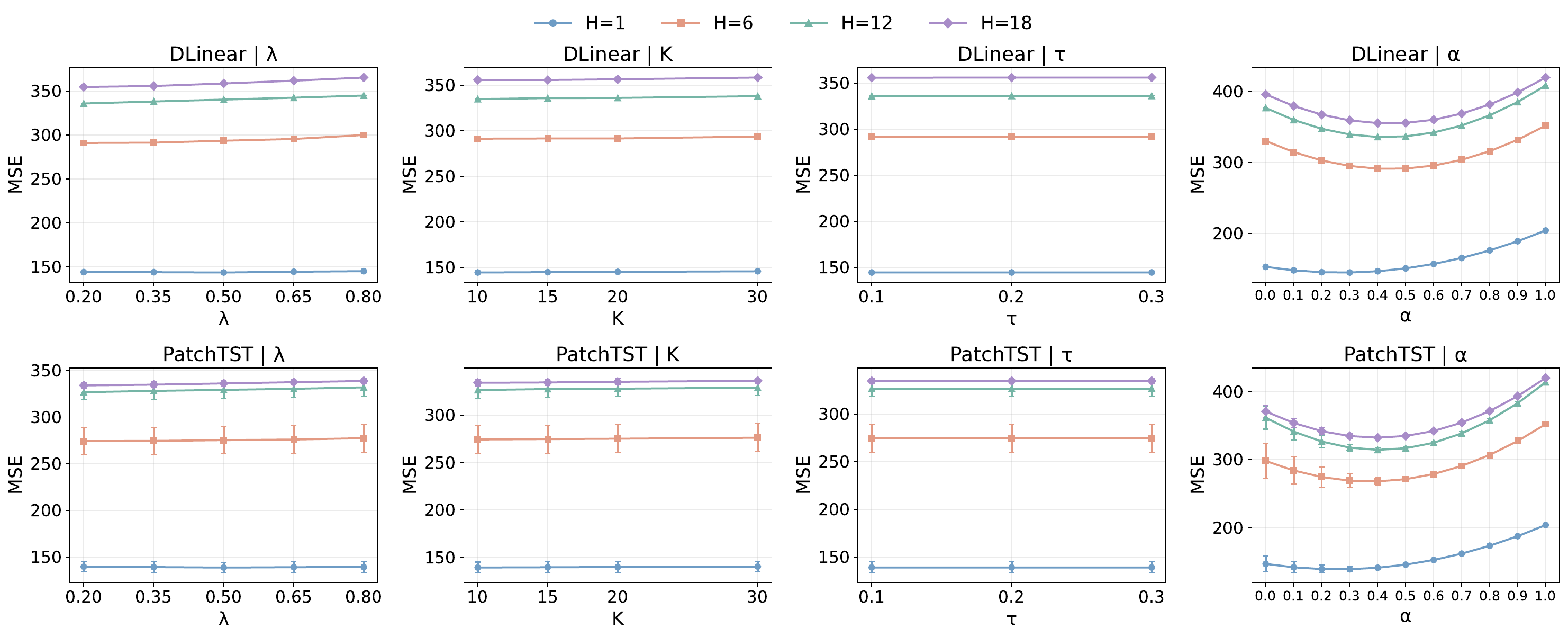}
    \caption{
    Parameter sensitivity on ASU Building 10 using DLinear and
    PatchTST. Each curve reports the mean MSE over three random seeds
    for forecasting horizons $H\in\{1,6,12,18\}$, with error bars
    denoting one standard deviation. Each parameter is varied
    independently while the remaining configuration is fixed to the
    values selected by validation. Lower MSE is better.
    }
    \label{fig:bn040_parameter_sensitivity}
\end{figure}

We investigate the sensitivity of the retrieval-correction module
using DLinear and PatchTST on Building 10 from the ASU dataset,
considering forecasting horizons $H\in\{1,6,12,18\}$.
Following a one-at-a-time sensitivity analysis, we vary
$\lambda\in\{0.20,0.35,0.50,0.65,0.80\}$,
$K\in\{10,15,20,30\}$,
$\tau\in\{0.10,0.20,0.30\}$, and
$\alpha\in\{0.0,0.1,0.2, 0.3, 0.4, 0.5, 0.6, 0.7, 0.8, 0.9,1.0\}$ independently.
For each model and forecasting horizon, all other parameters are fixed
to the configuration previously selected based on validation performance.

As shown in Fig.~\ref{fig:bn040_parameter_sensitivity}, the four
retrieval parameters exhibit different levels of sensitivity.
Overall, $\lambda$ and $K$ have relatively limited effects on
forecasting performance for both DLinear and PatchTST. For
$\lambda$, the MSE shows a mild increasing trend as its value
grows, particularly at longer forecasting horizons. Since a larger
$\lambda$ assigns greater weight to load-state similarity and less
weight to weather similarity, this observation suggests that
longer-horizon retrieval may benefit more from external weather
information than from relying predominantly on recent load
patterns. Similarly, varying $K$ from 10 to 30 results in only
minor changes in MSE across all forecasting horizons, indicating
that the method is not highly sensitive to the exact number of
retrieved neighbors within the evaluated range. The temperature parameter
$\tau$ is even less sensitive, with nearly overlapping curves over
$\tau\in\{0.10,0.20,0.30\}$ for both forecasting backbones and all
four horizons. These results indicate that the retrieval mechanism is
largely robust to the choice of $\tau$ within the evaluated range.

In contrast, $\alpha$ has a substantially stronger influence on
forecasting performance. Both models exhibit a consistent U-shaped
trend across all four horizons, with the lowest MSE occurring around
$\alpha=0.3$ for $H=1$ and around $\alpha=0.4$ for the longer
horizons. Performance deteriorates when $\alpha$ is either too small
or, more noticeably, too large, suggesting that effective correction
requires an appropriate balance between the original backbone forecast
and the retrieval-based adjustment. Notably, the favorable region
remains concentrated around $\alpha=0.4$ despite the different
error scales across forecasting horizons. Overall, the results show
that the method is relatively robust to $\lambda$ and $K$, particularly
insensitive to $\tau$, while $\alpha$ is the primary parameter
governing the strength of retrieval correction.

\subsection{Computational Cost}
\label{sec:computational_cost}

We evaluate the computational cost of the proposed retrieval
branch and correction module with MLP on UCF Building 30, using an input length of 24
and a forecasting horizon of $H=18$. Experiments are conducted
on one NVIDIA GH200 GPU with four CPU threads. The retrieval
parameters are fixed at $\lambda=0.2$, $K=10$, $\tau=0.1$, and
$\alpha=0.5$, with 18,353 entries stored in the training-only
retrieval memory. Inference timing excludes model initialization,
memory construction, and file I/O, assuming that the required
models, data, and retrieval memory are already loaded.

As shown in Table~\ref{tab:computational_cost}, building the
retrieval memory takes 5.668\,s and requires 28.775\,MiB of
storage. The retrieval and fusion process has a per-sample
latency of 6.135\,ms. For the full test set of 5,238 samples,
the total runtime of post-hoc fusion is $7.232$\,s,
indicating a modest computational cost for the proposed
retrieval correction.

\begin{table*}[t]
\centering
\caption{Computational cost of the retrieval-correction mechanism with MLP on
UCF Building 30 at forecasting horizon $H=18$.}
\label{tab:computational_cost}
\small
\setlength{\tabcolsep}{9pt}
\begin{tabular}{ccccc}
\toprule
Memory build time
&
Memory size
&
Per-sample latency
&
Full-test runtime
&
Test samples
\\
\midrule
5.668\,s
&
28.775\,MiB
&
6.135\,ms
&
7.232\,s
&
5,238
\\
\bottomrule
\end{tabular}
\end{table*}

%% file: sections/05-retrieval-correction-analysis.tex
\section{Case Study}
\label{sec:case_study}

WeL-RAC can offer case-based interpretability by explicitly linking
forecast corrections to retrieved historical evidence. This
section examines selected representative success and failure cases
to illustrate how that evidence influences the final prediction.
By comparing the initial forecasts, retrieved historical
continuations, and observed loads, we further analyse how retrieval
compensates for forecasting errors and why incompatible historical
load levels can instead lead to harmful corrections.

\subsection{Successful case studies}
\begin{figure}
    \centering
    \includegraphics[width=\textwidth]{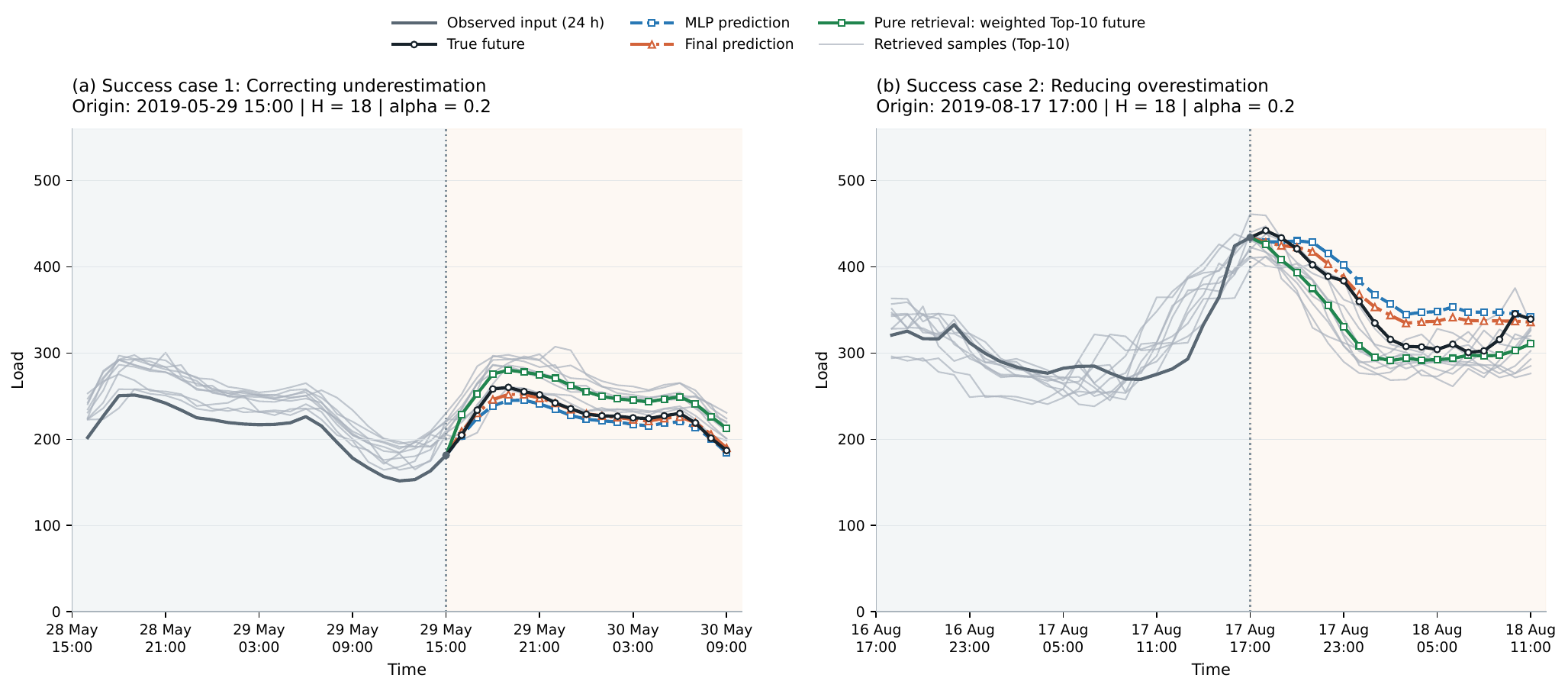}
    \caption{
       Two representative success cases of retrieval-assisted MLP forecasting on ASU Building 10.
    Both examples use an 18-hour forecasting horizon, a 24-hour input window,
    and seed 2021.
    (a) Retrieval corrects an underestimation case with forecast origin
    29 May 2019 at 15:00.
    (b) Retrieval reduces overestimation with forecast origin
    17 August 2019 at 17:00.
    The dark-gray line shows the preceding 24 hourly observations, and the
black line with circle markers shows the true future.
The blue dashed, green solid, and orange dash-dotted curves denote the
MLP prediction, weighted Top-10 retrieval prediction, and final prediction,
respectively.
Light-gray curves show the input histories and subsequent load trajectories
of the $K$ retrieved training samples. Historical inputs and continuations are training-period observations,
    aligned to the query by their respective origins without changing load values.
    }
    \label{fig:mlp_h18_success}
\end{figure}

Fig.~\ref{fig:mlp_h18_success} illustrates how retrieval complements
the MLP forecast in two selected 18-hour prediction windows for
ASU building 10. In Fig.~\ref{fig:mlp_h18_success}(a), the MLP
underestimates the observed load throughout the forecast horizon,
whereas the weighted historical continuation lies above the
observations. Combining these predictions with the fixed,
validation-selected retrieval weight $\alpha=0.2$ reduces MSE
from $81.33$ to $18.86$ ($76.81\%$), showing that
its contribution arises from compensating for the direction of
the forecasting error.
Fig.~\ref{fig:mlp_h18_success}(b) exhibits the reverse correction:
following the rise in the observed input, the MLP overestimates
much of the subsequent declining load trajectory, while the
retrieved continuation remains below the observations.
The same fusion weight shifts the forecast downward, reducing
MSE from $961.79$ to $481.19$ ($49.97\%$). Across both cases,
the historical trajectories therefore provide a correction
whose direction complements the MLP error, while the dominant
MLP contribution limits the influence of their load-level
mismatch. These examples demonstrate the value of error
complementarity under a fixed fusion rule; the benefit does
not require the retrieved continuation to be individually
more accurate than the MLP. These two success cases are selected for illustration; therefore, the reported improvements reflect these specific examples rather than the average performance over the full test set.

\subsection{Failure case studies} % need together with initial analysis by codex
\label{sec:failure_analysis}

\begin{figure}
    \centering
    \includegraphics[width=\textwidth]{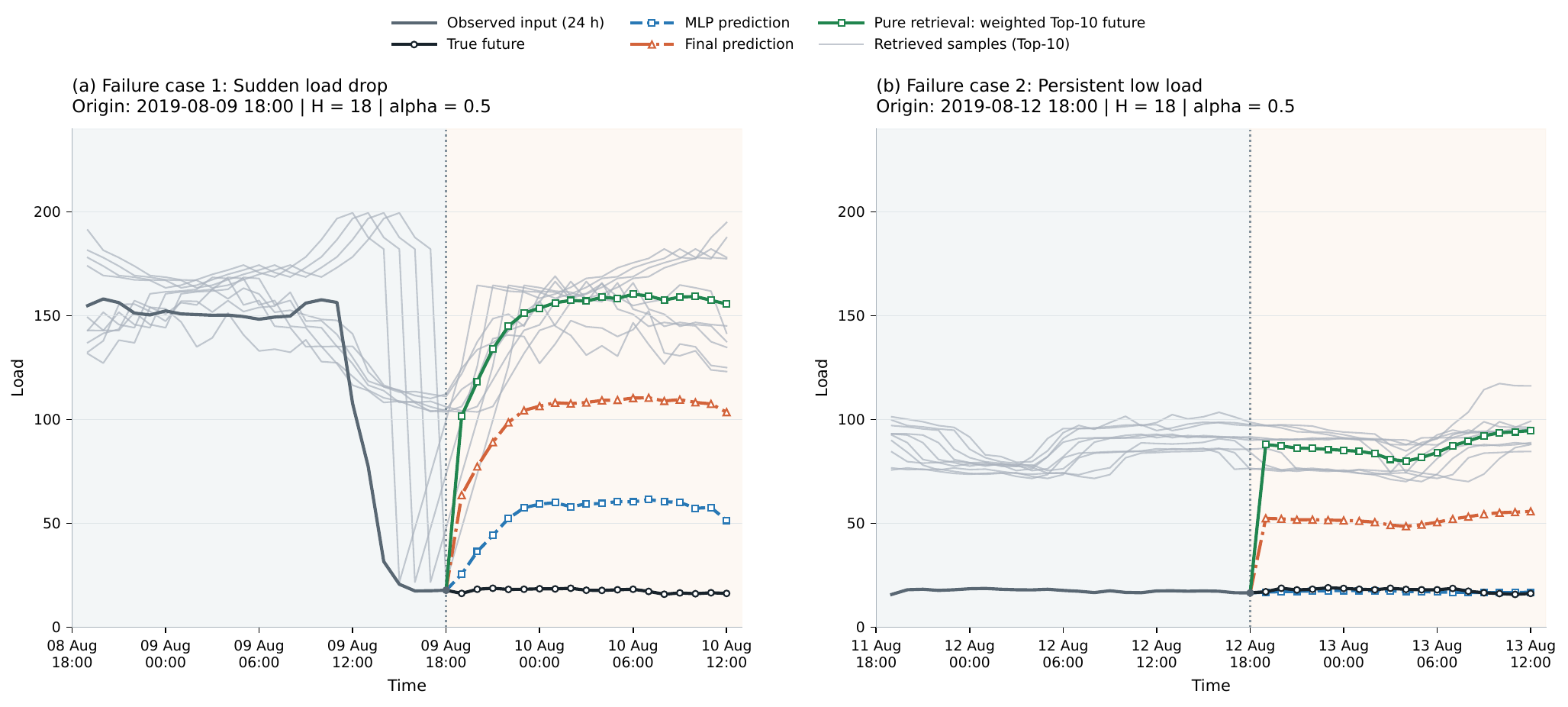}
    \caption{
    Two representative failure cases of retrieval-assisted MLP forecasting on UCF Building 30.
Both examples use an 18-hour forecasting horizon and seed 2021.
(a) A sudden load-drop case with forecast origin 9 August 2019 at 18:00.
(b) A persistent low-load case with forecast origin 12 August 2019 at 18:00.
The dark-gray line shows the preceding 24 hourly observations, and the
black circles indicate the true future. The blue dashed, green solid,
and orange dash-dotted curves represent the MLP prediction, weighted
Top-10 retrieval prediction, and final prediction, respectively.
Light-gray curves show the input histories and subsequent load trajectories
of the ten retrieved training samples. Historical samples are aligned by
their forecast origins and retain their original load levels.}
    \label{fig:mlp_failure_cases}
\end{figure}

Retrieval can degrade forecasting accuracy when the retrieved historical
continuations are incompatible with the current load condition. For the
MLP forecaster on Building~30 in the UCF dataset at $H=18$~h,
retrieval fusion increases the three-seed mean test MSE from $197.03$ to
$228.32$, a deterioration of $15.88\%$. Fig.~\ref{fig:mlp_failure_cases}
examines two forecast origins within the same August 2019 low-load episode
using seed~2021. Both cases retain the original validation-selected
configuration, including the 24-hour input, Top-10 retrieval, and fusion
weight $\alpha=0.5$. These windows are selected post hoc to diagnose the
failure mechanism; they are not used to revise the reported configurations.

Fig.~\ref{fig:mlp_failure_cases}(a) shows a sharp reduction from
approximately $150$--$160$ to below $20$, followed by an
18-hour period with a mean load of $17.60$. This sustained low-load
condition is absent from the validation period used for hyperparameter
selection: validation contains no observations below approximately $78.82$. The test episode therefore
exposes a validation-to-test distribution shift in load level and
persistence. Hyperparameters selected on validation data had no
opportunity to be assessed against this abrupt transition into a
sustained extreme-low-load state. The predictions in Fig.~\ref{fig:mlp_failure_cases}(a) illustrate the
consequence of applying the selected configuration under this shift.
The MLP anticipates a partial recovery, with a mean prediction of
$54.57$, while the retrieved historical continuations favor a
stronger recovery and yield a weighted mean of $150.06$. The
fixed fusion weight of $0.5$ raises the final mean prediction to
$102.31$, increasing the window MSE from $1455.97$ to
$7333.66$. The retrieved recovery pattern is
inappropriate because the actual load remains low throughout the
forecast horizon. Together, the validation coverage and the plotted
trajectories indicate a limitation in transferring the selected
retrieval and fusion settings to an unseen load condition.

In Fig.~\ref{fig:mlp_failure_cases}(b), the recent input has already
stabilized at a low level. The query and several retrieved samples
exhibit qualitatively similar, relatively stable temporal profiles,
but their absolute load levels differ substantially. The query input
averages $17.57$~kWh, whereas the retrieved histories fluctuate around
much higher levels. Their weighted future averages $86.98$~kWh,
compared with the observed future mean of $17.82$~kWh. Here, the
prominent mismatch concerns scale: selecting broadly similar shape does not ensure an appropriate absolute load level.

This distinction is supported by the logged neighbor similarities.
The retrieval-weighted load-state cosine similarity is $0.980$, and
the corresponding weather similarity is $0.988$. These scores describe
similarity in the retrieval representations, rather than equality of
the raw load trajectories. Although the state representation includes
recent levels and trends, the current standardized and normalized
cosine matching does not sufficiently constrain absolute-load
compatibility in this case. The MLP itself predicts a mean of
$17.18$~kWh and achieves a window MSE of $0.96$.
Fusion with the higher-level retrieval output shifts the final mean
to $52.08$~kWh and increases MSE to $1181.92$.
Thus, retrieval can degrade an accurate MLP forecast even when the
current input no longer contains a sharp transition.

\subsection{Discussion and Implications}
\label{subsec:case_discussion}

The above cases highlight an inherent interpretability advantage of
WeL-RAC. Because the correction is explicitly derived from retrieved
historical trajectories, the source and direction of each adjustment
can be inspected directly. In the successful cases, the retrieved
future does not necessarily need to outperform the base forecast on
its own; instead, it is useful when the historical evidence provides
a correction direction that complements the forecasting error. This
supports the role of retrieval as interpretable auxiliary evidence
rather than as a replacement for the original forecaster. In contrast,
the failure cases make it possible to trace harmful corrections to
specific retrieval mismatches, such as unseen operating conditions or
retrieved trajectories with incompatible absolute load levels.

These observations also suggest several ways to improve retrieval
reliability. First, validation should cover more diverse operating
regimes, for example through rolling validation or targeted stress
tests, so that retrieval and fusion parameters are not selected only
under typical conditions. Second, neighbour selection could explicitly
consider load-level compatibility in addition to pattern similarity,
using information such as the recent mean load, last observed value,
and variability to scale or rerank retrieved candidates. Third, the
fusion weight could be made reliability-aware, reducing the influence
of retrieval when the retrieved evidence is poorly matched to the
current condition or strongly inconsistent with the base prediction.
Such a mechanism could provide a conservative fallback to the original
forecaster when retrieval is unreliable. Adaptive memory updates may
further improve coverage as building operating conditions evolve over
time.

% These implications are derived from diagnostic case studies rather
% than from separately validated reliability mechanisms. The selected
% examples illustrate identifiable success and failure modes, but do not
% establish how frequently they occur across buildings. Future work
% should therefore evaluate these mechanisms systematically across
% broader operating conditions and held-out periods.

%% file: sections/06-conclusion.tex
\section{Conclusion}
\label{sec:conclusion}

This study presented WeL-RAC, a model-agnostic framework that
complements data-driven building load forecasters with
explicitly retrieved historical evidence. By jointly matching
recent load states and horizon-aligned weather summaries, the
framework retrieves observed historical continuations and fuses
them with initial predictions without changing the base
forecaster's architecture or training objective. Experiments
across 31 building cases, eight forecasting backbones, and four
prediction horizons demonstrated lower MSE and MAE and higher
$R^2$ for all 24 combinations of backbone and dataset at the
dataset level. After averaging across backbones and horizons,
MSE and MAE improved for 30 and 28 of the 31 building cases,
respectively. Ablations identified load-state matching as the
primary retrieval signal and weather summaries as complementary
context, while revealing that compact state representations
were more effective on ASU and UCF, whereas raw load sequences
performed better on AlphaBuilding. Success and failure case
studies further illustrated the framework's case-based
interpretability by tracing beneficial error compensation and
harmful load-level mismatches to the retrieved evidence.

Nevertheless, the retrieval memory and validation-selected fusion
weight remain fixed during testing, and high similarity in the
normalised retrieval space does not guarantee compatibility in
absolute load levels. Consequently, retrieval can degrade an
otherwise accurate forecast under operating conditions not
represented during validation. The reliance on predefined
weather summaries and the use of oracle weather in AlphaBuilding
also warrant further investigation of sensitivity to
weather-forecast errors and the transferability of the summary
rules. Moreover, the evaluation is restricted to offline,
single-building point forecasting on selected synthetic scenarios
and two campus datasets, leaving broader operational applicability
to be established. Future work should investigate load-level
compatibility checks, reliability-aware fusion, and online memory
updates, alongside comparisons with alternative retrieval methods
and evaluation across additional building types and climates.
Prospective deployment and uncertainty-aware evaluation are
also needed to determine whether the observed forecasting gains
translate into reliable building-management benefits.